\documentclass[
 reprint,
superscriptaddress,
 amsmath,amssymb,
 aps,
]{revtex4-2}

\usepackage{graphicx}
\usepackage{dcolumn}
\usepackage{bm}

\begin{document}


\title{Stability and electronic properties of CsK$_2$Sb surface facets}

\author{Richard Schier}
\affiliation{Carl von Ossietzky Universität Oldenburg, Physics Department, D-26129 Oldenburg, Germany}
\affiliation{Humboldt-Universit\"{a}t zu Berlin, Physics Department and IRIS Adlershof, D-12489 Berlin}
\author{Holger-Dietrich Sa{\ss}nick}
\affiliation{Carl von Ossietzky Universität Oldenburg, Physics Department, D-26129 Oldenburg, Germany}
\author{Caterina Cocchi}
\email{caterina.cocchi@uni-oldenburg.de}
\affiliation{Carl von Ossietzky Universität Oldenburg, Physics Department, D-26129 Oldenburg, Germany}
\affiliation{Humboldt-Universit\"{a}t zu Berlin, Physics Department and IRIS Adlershof, D-12489 Berlin}

\date{\today}

\begin{abstract}
First-principles methods have recently established themselves in the field of photocathode research to provide microscopic, quantum-mechanical characterization of relevant materials for electron sources. While most of the existing studies are focused on bulk crystals, the knowledge of surface properties is essential to assess the photoemission performance of the samples. In the framework of density-functional theory, we investigate stability and electronic properties of surface slabs of CsK$_2$Sb, an emerging semiconducting photocathode material for particle accelerators. Considering surfaces with Miller indices 0 and 1, and accounting for all possible terminations, we find that, at the interface with vacuum, the atomic layers may rearrange considerably to minimize the electrostatic repulsion between neighboring alkali species. From the analysis of the surface energy as a function of the chemical potential, we find a striking preference for surfaces oriented along the (111) direction. Yet, at large and intermediate concentrations of Cs and K, respectively, (100) and (110) slabs are energetically most favorable. The considered surfaces exhibit either semiconducting or metallic character. Of the former kind is the most stable (111) slab which has a band gap of about 1.3~eV, in excellent agreement with experimental values for CsK$_2$Sb samples. Metallic surfaces have lower work function, on the order of 2.5~eV, in line with the emission threshold measured for CsK$_2$Sb photocathodes. All in all these results contribute to the fundamental understanding of the microscopic properties of CsK$_2$Sb in particular and of multi-alkali antimonides in general, and represent an useful complement to the ongoing experimental efforts in the characterization of this emerging class of photocathode materials.
\end{abstract}

\maketitle


\section{Introduction}
The study of multi-alkali antimonides as novel materials for electron sources has been stimulated by the need to optimize the photoemitting performance of photocathodes for an efficient generation of ultrabright electron beams~\cite{musu+18nimpra,panu+21nimpra}.
Extended experimental research in this field~\cite{vecc+11apl,schu+13aplm,schu+16jap,feng+17jap,ding+17jap,gaow+17aplm,schm+18prab} has been complemented in the last few years by a number of \textit{ab initio} studies which have substantially contributed to the microscopic understanding of these materials~\cite{murt+16bms,kala+10jpcs,kala+10jpcs1,cocc+18jpcm,cocc+19sr,cocc20pssrrl,amad+21jpcm,cocc-sass21micromachines,khan+21ijer,sass-cocc21es,shu+21optik}.
Analysis of formation energies~\cite{cocc-sass21micromachines}, electronic properties~\cite{kala+10jpcs,kala+10jpcs1,cocc+18jpcm,cocc+19sr,amad+21jpcm,sass-cocc21es,cocc-sass21micromachines}, optical absorption~\cite{kala+10jpcs,kala+10jpcs1,cocc+18jpcm,amad+21jpcm,cocc-sass21micromachines,khan+21ijer}, and even core-level spectroscopy~\cite{cocc+18jpcm,cocc+19sr,cocc20pssrrl,cocc-sass21micromachines} have brought unprecedented insight into the relationships among structure, composition, and light-matter interactions in this class of compounds. 
This knowledge is essential to support the optimization of the photocathode growth~\cite{schm+18prab,yama+18ami,gald+19quantum} and the experimental characterization of resulting samples~\cite{xie+17jpd,gaow+17aplm,gevo+18prab,gald+21apl}.

The photoemission process of cathode materials is commonly interpreted and predicted through the so-called three-step model~\cite{berg-spic64pr}, which is usually fed by empirical parameters.
This approach facilitates the interpretations of experimental findings but limits the applicability of the model only to available samples. 
Recently, \textit{ab initio} studies on materials with favorable performance as photocathodes~\cite{anto+20prb,nang+21prb} have provided suitable data for the three-step model, thereby substantially enhancing its predictive power.
However, to quantitatively predict the photoemission yield of a material, the sole knowledge of its bulk properties is insufficient, as relevant quantities such as surface stability and
work function cannot be predicted from the bulk.
Likewise, atomic termination and relaxation at the interface with vacuum are key characteristics of the surfaces that can play a significant role in the physico-chemical properties of the photocathodes~\cite{mamu+17prab,xie+17jpd,dai+20,panu+21nimpra}.
The most recent advances in the growth of alkali-antimonide single crystals with well-defined orientation with respect to a substrate~\cite{parz+22prl} further motivate a systematic analysis of the surface properties of these materials. 

In this paper, we present a first-principles study based on density-functional theory (DFT) on the energetic stability and the electronic properties of CsK$_2$Sb surface facets with Miller indices 0 and 1.
In constructing the slabs, we account for all possible terminations arising from the ternary composition of this material, resulting in a pool of seven systems. 
After relaxation, which leads in some cases to a substantial rearrangement of the outermost atomic layers with respect to their configuration in the bulk, we evaluate the surface stability as a function of the chemical potential of the involved species and obtain a clear preference for the (111) direction.
From the analysis of the electronic structure, conducted with a state-of-the-art approximation for the exchange-correlation potential, we find three semiconducting surfaces with band gaps ranging from 0.8~eV to approximately 1.3~eV, three metallic slabs with work functions between 2.3 and about 3~eV, and one system with partially unoccupied valence states at 0~K which is expected, however, to be influenced by increasing temperature.
From the projected density of states we identify correlations between the character the electronic levels crossing the Fermi energy in the metallic surfaces and the atomic species which exceed at the interface with vacuum. 
These are useful indications to better understand the photoemission yield of these surfaces and to gain insight into their reactivity.

\section{Methods}

\subsection{Computational details}
All results presented in this work are obtained from DFT~\cite{hohe-kohn64pr,kohn-sham65pr}, implemented in the Gaussian and plane-wave formalism provided by the CP2K package~\cite{kueh+20jcp}.
The short-range-triple-$\zeta$ MOLOPT basis set~\cite{molopt2007} is adopted, with the atomic cores represented by the Godecker-Teter-Hutter (GTH) pseudopotentials~\cite{GTH1996, GTH1998, GTH2005}.
The plane-wave expansion of the density is truncated at the kinetic energy threshold of 600~Ry, and a homogeneous \textbf{k}-mesh with 12 points in all periodic directions adopted to sample the Brillouin zone of bulk and surfaces.
The generalized-gradient approximation in the Perdew-Burke-Ernzerhof (PBE)~\cite{pbe} parameterization is employed to relax the systems and to evaluate their energetic stability, while the strongly constrained and appropriately normed (SCAN) functional~\cite{sun+15prl} is adopted to calculate the electronic properties -- band structure, projected density of states (PDOS), and work function -- of the relaxed structures in single-point geometry calculations.
The choice of SCAN is motivated by the results of recent studies demonstrating the superior trade-off between accuracy and computational costs of this functional to predict the electronic properties of cesium antimonides and tellurides~\cite{sass-cocc21es,cocc-sass21micromachines,sass-cocc22jcp}.

\subsection{Construction of the slabs}
The slabs are generated using the Pymatgen library~\cite{sun-cede13ss} starting from the face-centered-cubic structure of bulk CsK$_2$Sb taken from Materials Project~\cite{jain+13aplm} -- entry number mp-581024~\cite{CsK2Sb} -- and further optimized until the interatomic forces are lower than 0.01~eV/\AA{}.
The slab thickness has been optimized in order to avoid unphysical effects due to spurious nanoconfinement: convergence tests and further details about the size of the systems are provided in the Supplemental Material (SM)~\cite{sm}, Fig.~S1 and Table~SI.
Periodic boundary conditions are applied only in two directions and 14~\AA{} of vacuum are included in the non-periodic one (see Fig.~S2~\cite{sm}).
For the calculation of the PDOS of the slabs, 3$\times$3 supercells have been constructed and simulated with $\Gamma$-point-only calculations as imposed by the current implementation of this feature in CP2K.
This procedure has been extensively used and tested for various systems (see, \textit{e.g.}, Refs.~\cite{ning+16ijhe,ansa+17prm}) even against experimental data~\cite{lind+1jpcc}. 
In the evaluation of the surface energies, we have considered the most stable phases of the elemental bulk crystals associated to the atomic constituents of CsK$_2$Sb, namely Cs, K, and Sb, taken from Materials Project~\cite{jain+13aplm}, entry numbers mp-1 (Cs)~\cite{Cs}, mp-58 (K)~\cite{K} and mp-104 (Sb)~\cite{Sb}. 
For consistency, these structures have been optimized with the same procedure and computational parameters listed above.

\section{Results and discussion}
\subsection{Structural properties}
\label{sec:structure}

\begin{figure}
	\centering
	\includegraphics[width=0.5\textwidth]{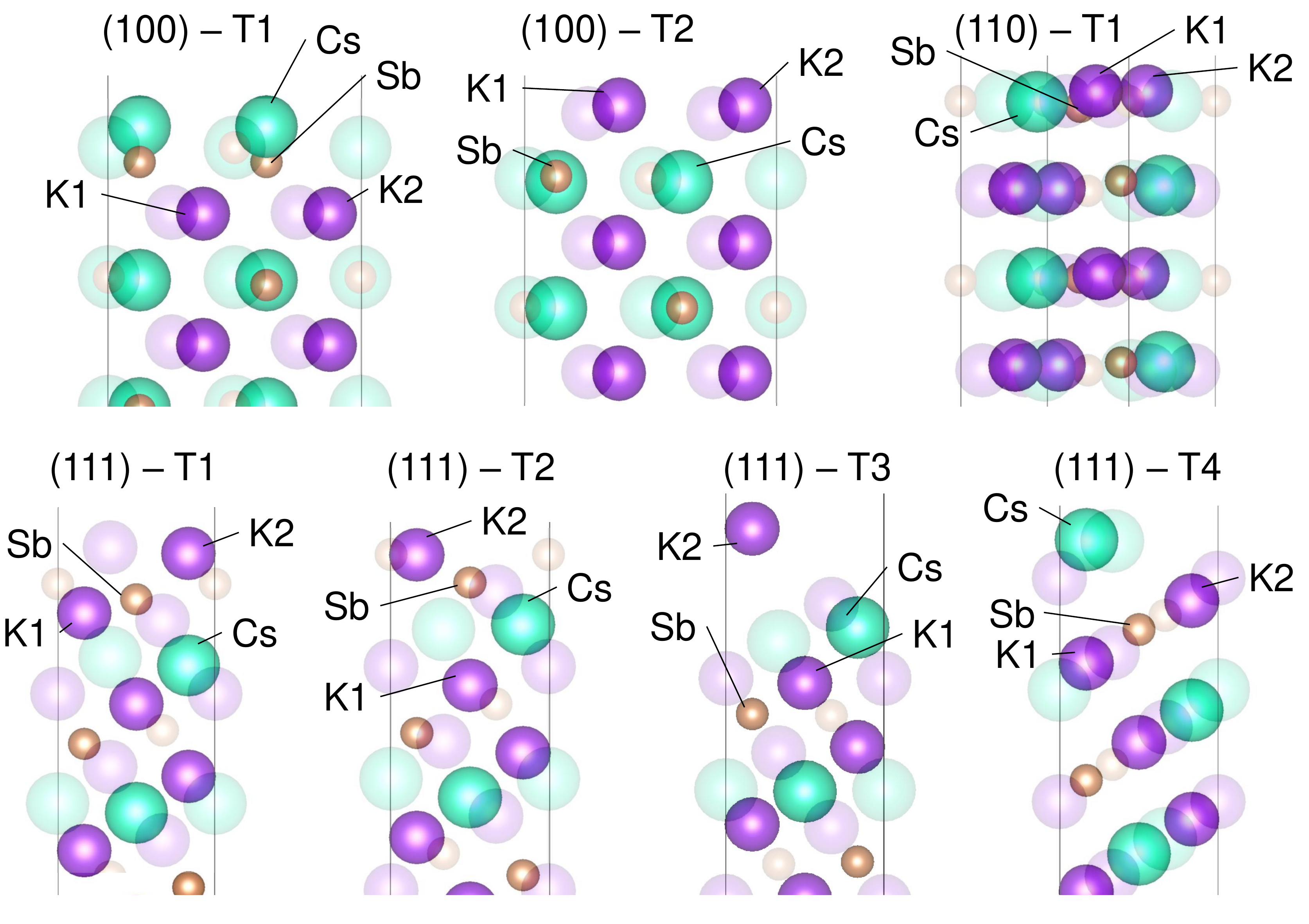}
\caption{Schematic view of the outermost atomic layers of the relaxed surface slabs considered in this work (bright spheres) compared with their initial structures constructed from the bulk (faded spheres). 
}
	\label{fig:relaxation}
\end{figure}

The construction of surface slabs for a ternary material like CsK$_2$Sb brings about an intrinsic complexity that needs to be carefully addressed.
Even when considering surfaces with low Miller indices 0 and 1, a relatively large number of terminations emerges, see Fig.~\ref{fig:relaxation}. 
For the (100) surface, two terminations are available: one with Cs and Sb atoms on the top layer, labeled T1, and one including only K atoms (T2).
The (110) slab admits only one termination containing all three species in the surface layer, while for the (111) surface the situation is more intricate.
As shown in Fig.~\ref{fig:relaxation}, bottom panels, four terminations exist in this case: T1 and T3 are characterized by an excess of K atoms on the outermost layer, T2 by an excess of Sb, while T4 is terminated by Cs atoms.
Slabs cut along the (110) directions are nonpolar while (100) and (111) slabs form so-called \textit{Tasker type-3} surfaces~\cite{nogu00jpcm}.
It is worth noting that only the (110) slab has the same stoichiometry as the bulk material. 
In all the other surfaces, the stochiometry of the parent bulk has been lifted to symmetrize the two ends of the slab (see Fig.~S2~\cite{sm}), which are originally asymmetric.
This pre-processing is needed to ensure comparable results among all considered surfaces.
We checked that this procedure does not introduce artifacts in the electronic properties of the systems, see Fig.~S3 in SM~\cite{sm}.
From a crystallographic viewpoint, the (100) and (110) surfaces are described by orthorhombic unit cells while those cut in the (111) direction by hexagonal ones.
The corresponding lattice parameters are reported in the SM~\cite{sm}, Table~SII.

Upon relaxation of the slabs, the atomic positions in the outermost layers rearrange themselves with respect to the bulk (see Fig.~\ref{fig:relaxation}).
In the two slabs cut along the (100) direction, the relaxed structures do not depart dramatically from their starting points but variations are still non-negligible. 
In (100)-T1, the under-coordinated Sb atoms rearrange themselves towards the inner layers due to the attraction to the K atoms.
Corresponding K-Sb bond lengths are indeed reduced by about 5\% compared to the bulk (see Table~\ref{tab:distances}).
In return, Cs atoms move towards the vacuum in response to the repulsion with the K atomic layer (see Mulliken charges in Table~SV~\cite{sm}), giving rise to an expansion of the K-Cs distances by almost 14\%. In the whole atomic rearrangement at the interface with vacuum, K-K bonds remain unaltered with respect to the unrelaxed configuration, while Cs-Sb bonds are slightly expanded.
The situation is similar in the (100)-T2 slab: The uppermost layer only consists of under-coordinated K atoms which move outwards, being repelled by the underlying Cs atoms. This process gives rise to an increase in the K-Cs bonds by 6.6\% (see Table~\ref{tab:distances}). The other two species, Cs and Sb, remain close to their positions in the bulk, as confirmed by their unaltered bond length.
K-K distances are unchanged, too, while the K-Sb ones increase by less than 3\%.
These mild variations cannot be perceived in Fig.~\ref{fig:relaxation}, where the visualized displacements of the atoms in the relaxed surface with respect to the initial arrangement are related to the periodicity of the system in the lateral directions.

In the (110) surface, the main effect of structural optimization is the sizeable expansion of the K-K distance by 16.4\%, see Table~\ref{tab:distances}, as a consequence of the mutual repulsion of these atoms at the interface with vacuum (see Mulliken charges in Table~SV~\cite{sm}) and in absence of an anion sufficiently close by. 
The other bond lengths are either unchanged (Cs-Sb), or mildly enlarged, as for the K-Cs distances, or reduced (K-Sb separations, -3.9\%).
Similar to the (100)-T2 slab, these effects are not visible in Fig.~\ref{fig:relaxation}, where, instead, the lateral atomic displacements within the periodicity of the surface in two directions can be appreciated.

\begin{table*}
\caption{Distances between the four inequivalent atoms in the outermost layers of the considered slabs and relative variations with respect to the bulk values reported on the first row. The involved atoms are marked in Fig.~\ref{fig:relaxation}. }
\label{tab:distances}
\begin{ruledtabular}
\begin{tabular}{lllllll}
Distances [\AA{}] & Cs-Sb & K1-K2  & K1-Sb & K2-Sb & K1-Cs & K2-Cs \\
\colrule
Bulk & 4.39  & 4.39 & 3.80  & 3.80  & 3.80  & 3.80  \\ \hline
(100)-T1         & 4.56 (+3.9~\%)  & 4.39 (+0.0~\%) & 3.59 (-5.5~\%)  & 3.59 (-5.5~\%)  & 4.33 (+13.9~\%)  & 4.33 (+13.9~\%)  \\
(100)-T2         & 4.40 (+0.0~\%) & 4.39 (+0.0~\%) & 3.92 (+2.9~\%)  & 3.92 (+2.9~\%) & 4.05 (+6.6~\%) & 4.05 (+6.6~\%) \\
(110)-T1         & 4.40 (+0.0~\%)  & 5.11 (+16.4~\%) & 3.65 (-3.9~\%)  & 3.65 (-3.9~\%)  & 4.04 (+6.3~\%) & 4.04 (+6.3~\%) \\
(111)-T1         & 4.28 (-2.5~\%) & 4.16 (-5.2~\%) & 3.62 (-4.7~\%)  & 3.93 (+3.4~\%) & 4.03 (+6.1~\%) & 3.94 (+3.7~\%) \\
(111)-T2         & 3.88 (-11.6~\%) & 5.82 (+32.6~\%) & 3.61 (-5.0~\%) & 3.72 (-2.1~\%) & 4.34 (+14.2~\%) & 4.17 (+9.7~\%) \\
(111)-T3         & 4.68 (+6.7~\%) & 6.44 (+46.7~\%) & 3.74 (-1.6~\%) & 3.76 (-1.1~\%) & 4.07 (+7.1~\%) & 4.96 (+30.5~\%) \\
(111)-T4         & 4.73 (+7.7~\%) & 4.17 (-5.0~\%) & 3.72 (-2.1~\%) & 3.77 (-0.1~\%)  & 4.23 (+11.3~\%) & 4.16 (+9.5~\%) \\
\end{tabular}
\end{ruledtabular}
\end{table*}

For the surfaces cut along the (111) direction, the situation is more complex.
In contrast with their siblings along (100) and (110), in these slabs, 
an alkali atom of the bulk unit cell stick outs and leans toward the vacuum (see Fig.~\ref{fig:relaxation} and Fig.~S2~\cite{sm}).
This uneven starting point leads to substantial atomic rearrangements in the optimized geometries.
In the T1 termination, the outermost K and Sb atoms move inwards, leading to a reduction of the Cs-Sb and K-K distances by a few percentage points, and by an increase in the K-Cs separations by similar amounts in absolute values (see Table~\ref{tab:distances}).
Interestingly, in this slab, the distances between the two inequivalent K atoms and Sb behave oppositely: the one between K1 and Sb is reduced by less than 5\% while K2-Sb increases by 3.4\%.
Although these variations are not enormous, the overall rearrangement is clearly visible in Fig.~\ref{fig:relaxation}.

In (111)-T2 and -T3, changes are more dramatic.
In the relaxed structure of the former, K atoms rearrange themselves approaching the vacuum while Sb atoms are repelled from it.
In the resulting configuration shown in Fig.~\ref{fig:relaxation}, it is apparent that this displacement in the outermost surface layer is favored by the anion (Sb) being sandwiched between the two alkali species, as in the (111)-T1 slab.
As a result, the distance between Sb and both cations reduces upon relaxation (see Table~\ref{tab:distances}).
Since the initial arrangement with K and Cs atoms occupying neighboring lattice sites is clearly unfavorable, the distances between them and between the two inequivalent K atoms increase dramatically with respect to the initial configuration.

A similar scenario prior to relaxation appears in (111)-T3, too.
In the starting configuration, K and Cs occupy the outermost atomic layers, with Sb appearing only deeper down in the slab (see Fig.~\ref{fig:relaxation}).
In the optimized structure, where an electrostatically favorable rearrangement of cations and anions is not possible, the outermost layer of K atoms is largely displaced from its initial position and, consequently, from the neighboring Cs atoms underneath. 
This gives rise to an extension of almost 50\% of the K-K separation and of more than 30\% of the K2-Cs one (see Table~\ref{tab:distances}). 
The other interatomic distances increase by smaller amounts with the exception of K2-Sb which decreases by a mere 1.1\%.

Finally, in the (111)-T4 slab, where in the initial configuration the two alkali species occupy neighboring sites at the surface, relaxation induces the displacement of K atoms away from the vacuum, giving rise to the Cs termination shown in Fig.~\ref{fig:relaxation}.
This rearrangement leads to an increase of the interatomic distances involving Cs: +7.7\% for Cs-Sb and about +10\% on average for each Cs-K distance (see Table~\ref{tab:distances}). As a side effect of the attraction between both inequivalent K atoms and Sb, the mutual distance between them reduces by 5\%: as visible from Fig.~\ref{fig:relaxation}, this is mainly due to the decrease of the K1-Sb bond length with respect to the input structure.

All in all, from the analysis reported above, we can group the considered slabs according to the chemical nature of their termination. 
Except for the (110) facet which hosts all atomic species on the top layer, all the other surfaces exhibit at the interface with vacuum one layer of cations [(100)-T1 and -T2, (111)-T1 and -T2], two layers of cations [(111)-T4], and three layers of cations [(111)-T3].
As we will discuss extensively in Section~\ref{sec:electronic}, this characteristic has a crucial impact on the electronic properties of the slabs.

It is worth mentioning that the rearrangement of the atomic positions at the surface is accompanied by a delocalization of the charge density with respect to the inner atomic layers, which in turn reflect the electronic distribution in the bulk.
While in the periodic crystal, Cs and Sb atoms are connected by a bond of pronounced ionic character (see Table~SV~\cite{sm}, where the values of the Mulliken charges are reported), at the interface with vacuum, the absolute values of the partial charges decrease systematically. 
In those slabs where one alkali atom sticks out, such as (111)-T3 and -T4, we even notice a change of sign in the charge of K2 and a reduction of the magnitude of the partial charge on Cs ranging between 40$\%$ and 50$\%$.
This charge redistribution acts to compensate the extended bond lengths that are present in these surfaces (see Table~\ref{tab:distances}) and, overall, to stabilize these structures.

We conclude this subsection remarking that the considered models do not include defects nor complex surface reconstruction. 
While both aspects very likely occur experimentally and are therefore relevant for the complete understanding of the physics of CsK$_2$Sb surfaces, their study requires dedicated analysis that can be performed only once the idealized scenario presented in this work has been fully understood. 

\subsection{Surface stability}\label{sec:energetics}
In the next step, we examine the \textit{energetic stability} of the considered slabs. 
For this purpose, we performed total energy calculations at 0~K as enabled by the adopted DFT formalism.
An alternative path involving the calculation of phonon bands as a way to assess the \textit{dynamical stability} of the surfaces is left for future work entirely dedicated to the analysis of the vibrational properties of these materials which are relevant not only for the evaluation of the photoemission yield~\cite{anto+20prb,nang+21prb} but also for the thermoelectric response of the systems~\cite{yue+22prb}.

For this analysis, it is necessary to introduce some key quantities. First, 
we define the surface energy $\gamma$ as:
\begin{equation}
\gamma = \frac{1}{2A} (E_{slab} - n_{Cs} \cdot \mu_{Cs} - n_{K} \cdot \mu_{K} - n_{Sb} \cdot \mu_{Sb}),
\label{eq:gamma}
\end{equation}
where $A$ is the surface area and $E_{slab}$ is the total energy of the slab,
$n_j$ is the number of atoms of species $j$, and $\mu_j$ is the corresponding chemical potential.
The factor 2 in the denominator accounts for the fact that in our atomistic models, each system is characterized by two surfaces (see Fig.~S2~\cite{sm}).
Assuming
\begin{equation}
\mu_{Cs} + 2\mu_{K} + \mu_{Sb} = \mu_{CsK_2Sb} \approx E_{CsK_2Sb}^{bulk},
\label{eq:mu_approx}
\end{equation}
where $E_{CsK_2Sb}^{bulk}$ is the total energy of bulk CsK$_2$Sb, we can express the chemical potential of one of the atomic constituents, Sb, in terms of the total energy of the bulk and of the chemical potential associated to the other two elements, $\mu_{Sb} = E_{CsK_2Sb}^{bulk} - 2\mu_{K} - \mu_{Cs}$, thereby effectively reducing the dependence of $\gamma$ to only two variables: $\gamma (\mu_{Cs}, \mu_{K}, \mu_{Sb}) = \gamma (\mu_{Cs}, \mu_{K})$.
This way, we can rewrite Eq.~\eqref{eq:gamma} as
\begin{widetext}
\begin{align}
    \gamma & = \dfrac{ E_{slab} - n_{Cs} \cdot \mu_{Cs} - n_{K} \cdot \mu_{K} - n_{Sb} \cdot (E_{CsK_2Sb}^{bulk} - \mu_{Cs} - 2\mu_{K})}{2A} \\ \nonumber
    & = \dfrac{ E_{slab} - n_{Sb} \cdot E_{CsK_2Sb}^{bulk} +(n_{Sb} - n_{Cs} )\mu_{Cs} + (2 n_{Sb} - n_{K})\mu_{K}}{2A},
\end{align}
\end{widetext}
where the chemical potentials of each species are computed from DFT as the total energy per atom of the corresponding elemental crystal.
For stable surfaces, the following conditions hold:
\begin{align} \label{eq:mu-smaller}
\nonumber
& \mu_{Cs} < E_{Cs}^{at}, \\
& \mu_{K} < E_{K}^{at}, \\ 
\nonumber
& \mu_{Sb} < E_{Sb}^{at},
\end{align}
where $E_{X}^{at}$ is the total energy per atom of the elemental phase of species $X$.
We can complement these conditions with the lower boundaries for the chemical potentials taking advantage of Eq.~\eqref{eq:mu_approx}:
\begin{align} \label{eq:mu-larger}
&    E_{CsK_2Sb}^{bulk} - 2E_{K}^{at} - E_{Sb}^{at} < E_{CsK_2Sb}^{bulk} - 2\mu_{K} - \mu_{Sb} = \mu_{Cs}, \\ \nonumber
&  \dfrac{1}{2}(E_{CsK_2Sb}^{bulk} - E_{Cs}^{at} - E_{Sb}^{at}) < \dfrac{1}{2}(E_{CsK_2Sb}^{bulk} - \mu_{Cs} - \mu_{Sb}) = \mu_{K}
\end{align}
for Cs and K, respectively.
These equations can be solved knowing the number of atoms composing the slabs, the surface area, and the total energies of the slabs and of the bulk materials -- all quantities available as input or output of DFT calculations (see SM, ,Table~SIII and SIV~\cite{sm}).
Combining Eqs.~\eqref{eq:mu-smaller} and \eqref{eq:mu-larger}, we obtain the following conditions for the chemical potential of Cs
\begin{equation}
-549.795 \text{ eV} < \mu_{Cs} < -548.002 \text{ eV}, 
\end{equation}
and K
\begin{equation}
-769.876 \text{ eV} < \mu_K < -768.979 \text{ eV},
\end{equation}
respectively, which will be used to compute the stability of the considered surfaces without explicit dependence on $\mu_{Sb}$. This choice is consistent with the fact that the growth of CsK$_2$Sb samples entails the deposition of the two alkali species on an Sb. 

\subsubsection{Analysis of the surface energies}
\begin{figure}
	\centering
	\includegraphics[width=0.45\textwidth]{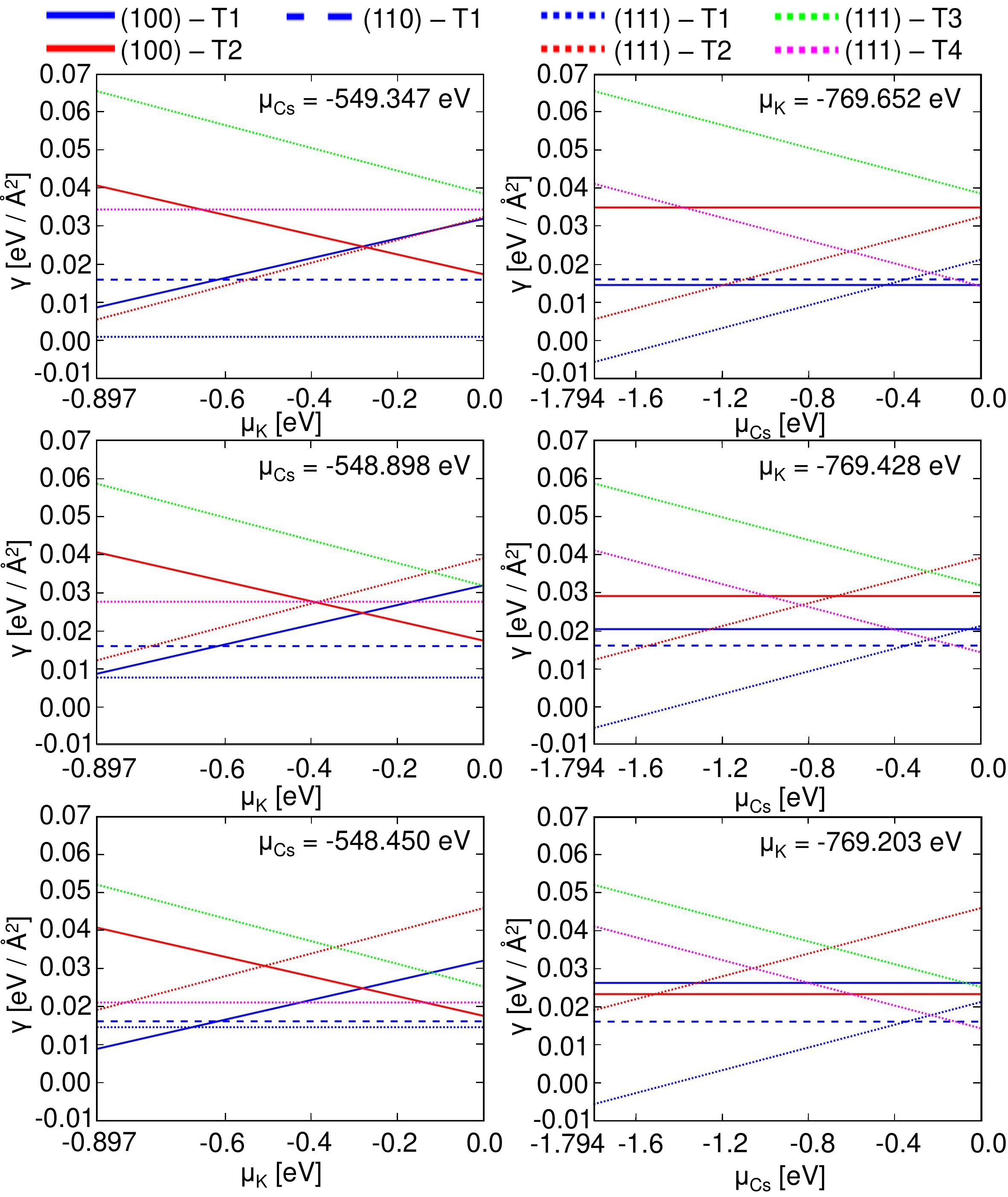}
	\caption{Surface energies of the considered CsK$_2$Sb slabs as a function of the chemical potential of K (left) and of Cs (right) for fixed values of $\mu_{Cs}$ and $\mu_K$, respectively. The $x$-axes are offset to $\mu_K = -768.979 \text{ eV}$ (left) and $\mu_{Cs} = -548.002 \text{ eV}$ (right).
	}
	\label{fig:stability}
\end{figure}

We can visualize the surface energy of each surface as a function of the chemical potential of the K and Cs, for fixed values of the other one, see Fig.~\ref{fig:stability}, left and right panels, respectively.
Generally, a lower surface energy indicates higher stability and, thus, a higher probability of that specific surface to be formed. 
When the chemical potential of Cs is kept fixed, regardless of its value, the probability of formation of (110)-T1, (111)-T1 and -T4 is constant with respect to $\mu_{K}$ (Fig.~\ref{fig:stability}, left).
On the other hand, both (100)-T2 and (111)-T3 become more stable at increasing values of $\mu_{K}$, in particular when $\mu_{Cs}$ increases, too.
The surface energies of (100)-T1 and (111)-T2 exhibit the opposite trend: their values raise (\textit{i.e.}, the formation of these slabs becomes less favorable) when $\mu_{K}$ increases.
It is worth noting that for these two systems, $\gamma$ has a different slope at varying (constant) value of $\mu_{Cs}$: for the most negative one considered in Fig.~\ref{fig:stability} (top-left panel), the two curves almost coincide, while for the other two values of $\mu_{Cs}$, they are almost parallel.

Moving now to the variation of the surface energy as a function of $\mu_{Cs}$ (Fig.~\ref{fig:stability}, right panels), we notice constant values for both slabs along the (100) direction and for the one along (110).
The formation of both (111)-T3 and -T4 slabs becomes more favorable with increasing $\mu_{Cs}$, while the opposite trend is exhibited by the surface energy of (111)-T1 and -T2.

All in all, (slightly) negative values of $\gamma$, which are indicative of the unusual preference of the material to crystallize in the surface rather than in the bulk, appear from our results only for (111)-T1 with the lowest values of $\mu_{Cs}$, irrespective of the magnitude of $\mu_{K}$.
This finding indicates this slab as most likely to form under the displayed conditions.
Furthermore, the general behavior of the formation energies shown in Fig.~\ref{fig:stability} suggests that the generation of K-rich/-poor environments can be used to nudge the system toward the formation of one surface or another one.
It should be noted that, except for the scenario of very low concentrations of Cs, differences in the formation energy are rather minor. 
Especially for the (110) surface, values reported in Fig.~\ref{fig:stability} are constant and very close to zero, which represents the bulk reference independent of the chemical potentials. This finding is consistent with the fact that this slab has the same stochiometry as the periodic crystal.
Notably, the trends displayed in Fig.~\ref{fig:stability} are in line with the behavior of the Cs$_3$Sb surfaces analyzed in Ref.~\cite{wang+17jpcc}.
In that work, it was shown that for a Cs-poor environment the (111) surface is the most stable one, while other surfaces become more favorable for Cs-rich conditions.

\begin{figure}
	\centering
	\includegraphics[width=0.5\textwidth]{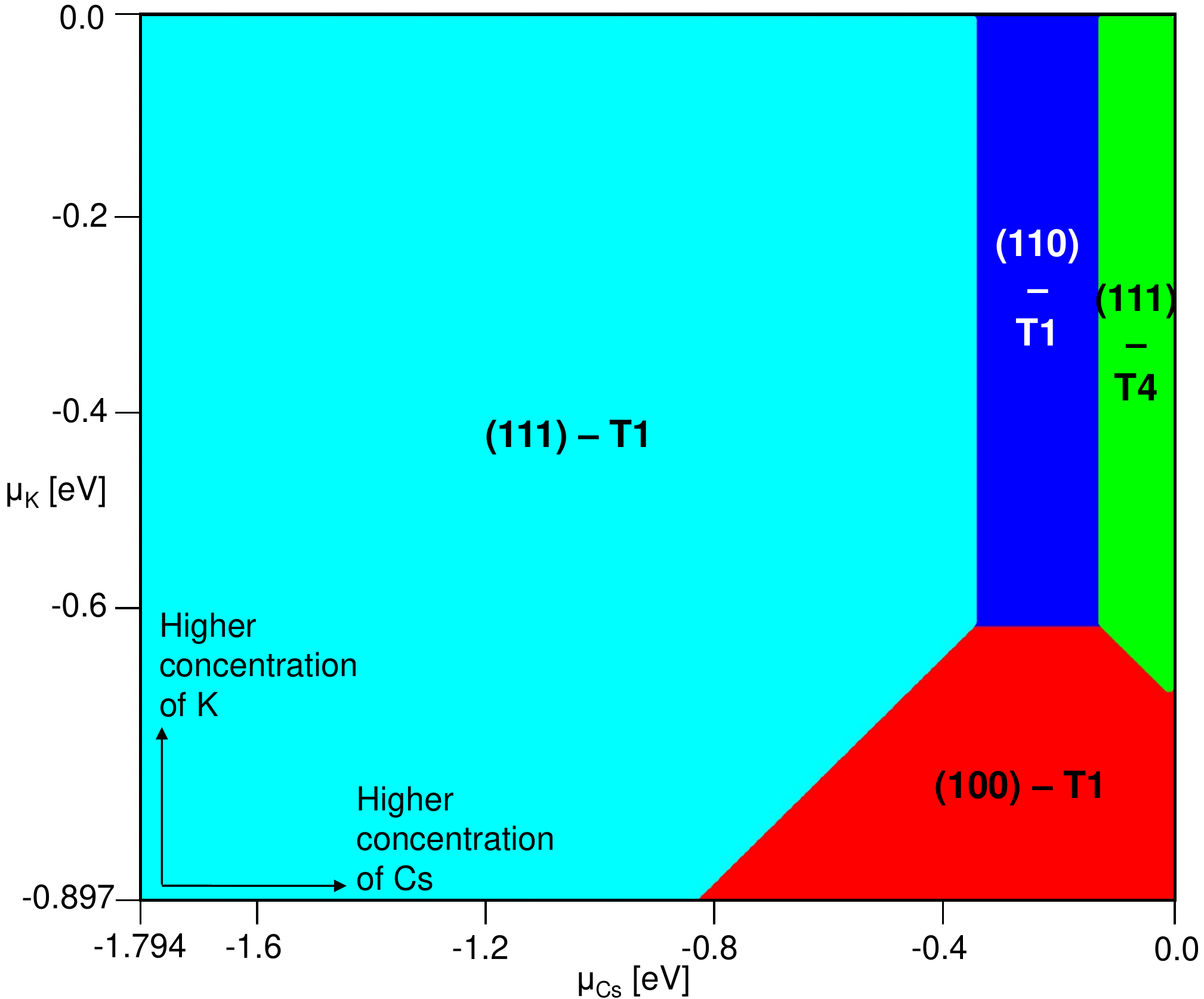}
	\caption{False color plot of the most stable surfaces as a function of $\mu_{Cs}$ and $\mu_{K}$. The $x$ and $y$ axes offset to $\mu_{Cs} = -548.002 \text{ eV}$ and $\mu_K = -768.979 \text{ eV}$, respectively.	}
	\label{fig:energetics}
\end{figure}

To summarize these results in a more intuitive manner, we show in Fig.~\ref{fig:energetics} the most stable slab predicted at the simultaneous values of the chemical potentials of Cs and K.
From this graph, it is evident that the (111)-T1 surface is the one that forms most favorably at all values of $\mu_K$ and at low Cs concentrations. 
At increasing values of $\mu_{Cs}$, beyond -0.8~eV, other slabs are most likely to be generated, depending whether $\mu_K$ is simultaneously below or above -0.6~eV.
In the former case, the (100)-T1 facet has the largest probability to be formed; in the latter scenario, the (110)-T1 slab, which isostoichiometric with the bulk, is energetically favored.
Finally, at concomitantly high values of $\mu_K$ and $\mu_{Cs}$, (111)-T4 becomes the most favorable surface.

\subsection{Electronic properties}
\label{sec:electronic}

\subsubsection{Band structures}
\begin{figure*}
	\centering
	\includegraphics[width=0.95\textwidth]{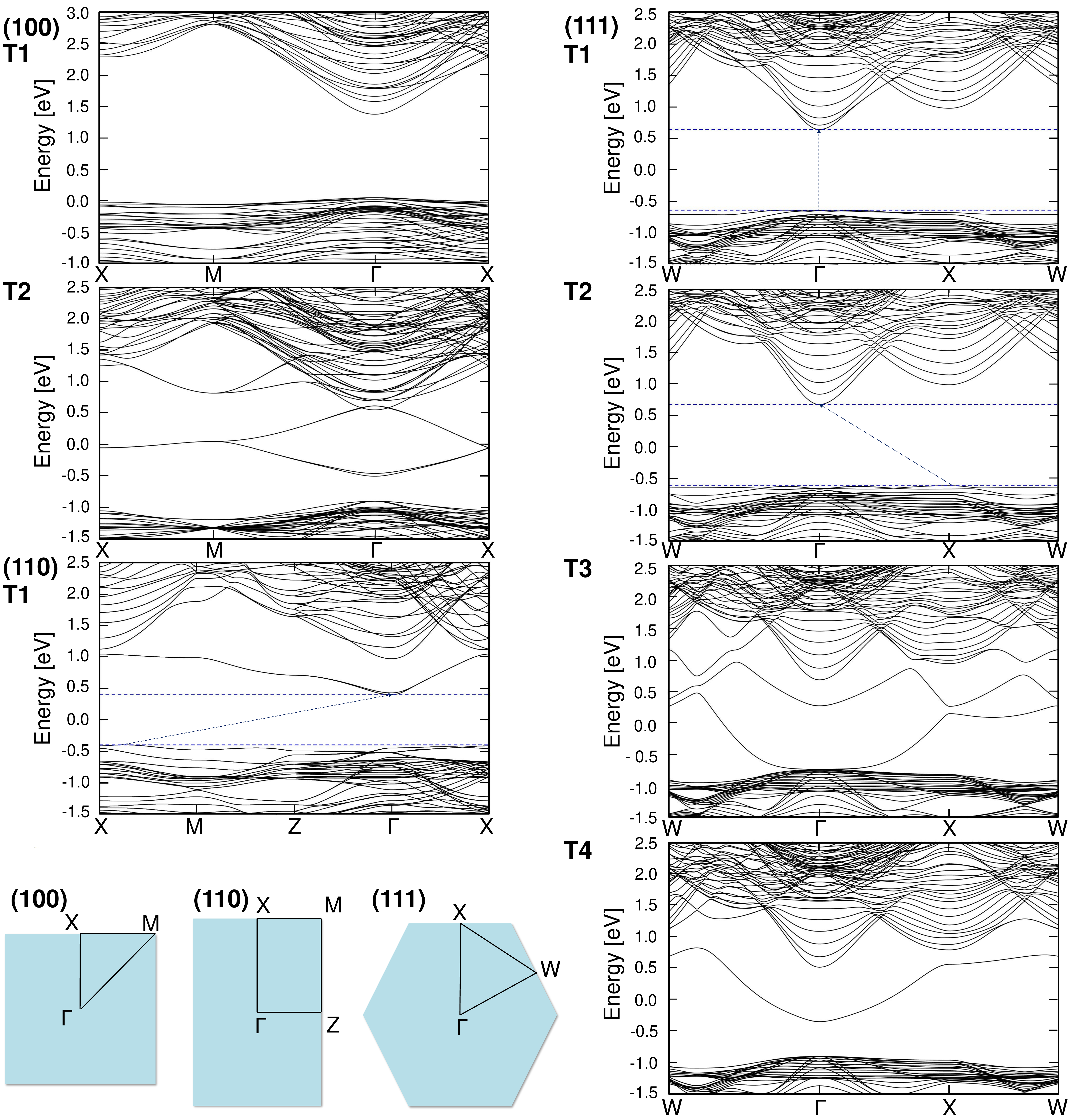}
	\caption{Band structures calculated with the SCAN functional of the seven CsK$_2$Sb slabs considered in this work offset with respect to the Fermi energy at 0~eV. In the semiconducting surfaces, the Fermi energy is set in the mid-gap and an arrow marks the fundamental gap between the valence band maximum and conduction band minimum highlighted by horizontal dashed lines. Bottom-left panel: Brillouin zones of the indicated surfaces including the paths connecting the high-symmetry points adopted in the band-structure plots.}
	\label{fig:bandstructure}
\end{figure*}

We continue our analysis with the characterization of the electronic properties of the considered CsK$_2$Sb surfaces starting from the inspection of their band structures (see Fig.~\ref{fig:bandstructure}).
Only three out of seven slabs are semiconducting like the parent bulk.
Among them we find the isostoichiometric facet (110) and the (111) slabs with T1 and T2 terminations. 
Both (110)-T1 and (111)-T2 exhibit a band gap between X and $\Gamma$ of 0.80~eV and 1.29~eV, respectively (see Table~\ref{tab:wf_ip}).
In (111)-T1, the band gap as large as 1.28~eV is direct at $\Gamma$ as in the bulk crystal~\cite{kala+10jpcs,murt+16bms,cocc+18jpcm,cocc+19sr}.
It is worth noting that even in (110)-T1 and (111)-T2, where the fundamental band gap is indirect, the optical gap at $\Gamma$ is only a few tens of meV larger (see Table~\ref{tab:wf_ip}).
The similarity between indirect and direct gaps in these systems is due to the very low dispersion of the highest valence bands, in line with the $p$-character of these states inherited from the bulk~\cite{cocc+18jpcm,cocc+19sr}.
Moreover, while in all these three semiconducting surfaces the lowest unoccupied state is found at $\Gamma$, in the (110)-T1 slab, it is energetically separated by several hundreds of meV from the next electronic levels in the Brillouin-zone center. 
In contrast, in both (111) surfaces, the conduction band minimum belongs to the lowest of a manifold of bands.
In all cases, the fundamental gap of the semiconducting slabs is reduced with respect to the corresponding value computed for the bulk at the same level of theory (SCAN functional)~\cite{cocc-sass21micromachines}.
Interestingly, the result obtained for the (111)-T1 surface, which is overall the most likely to form (see Fig.~\ref{fig:energetics}), is in excellent agreement with the experimental band gap of 1.2~eV reported for CsK$_2$Sb~\cite{ghos-varm78jap}.

\begin{table}
\caption{Work function $W$, ionization potential, $IP$, electronic band gap, $E_{gap}$, and optical band gap, $E_{gap}^{opt}$, of the considered surfaces calculated from DFT (SCAN functional). The band-gap values of the (100)-T1 slab are indicated in parenthesis since this structure is redicted to be metallic at 0~K.}
\label{tab:wf_ip}
\vspace{2mm}
\begin{tabular}{ccccc}
         & $W$ [eV] & $IP$ [eV] & $E_{gap}$ [eV] & $E_{gap}^{opt}$ [eV] \\ \hline \hline
(100) - T1 \;\;\; & - & 2.45 & (1.32) & (1.32)   \\ 
(100) - T2 \;\;\;& 2.33 & - & - & -  \\ 
(110) - T1 \;\;\;& - &  2.83 & 0.80 & 0.85  \\ 
(111) - T1 \;\;\;& - &  2.87 & 1.28 & 1.28   \\ 
(111) - T2 \;\;\;& - &  3.53 & 1.29 & 1.34   \\ 
(111) - T3 \;\;\;& 3.05 & - & - & -  \\ 
(111) - T4 \;\;\;& 2.45 & - & - & -  \\ \hline \hline
\end{tabular}
\end{table}

The remaining four slabs are metallic, in agreement with physical-chemical intuition based on their structure and composition.
In contrast with the semiconducting surfaces, which include Sb close to the interface with vacuum, all metallic systems exhibit an
excess of alkali atoms (either Cs or K) at the outermost layers (see Fig.~\ref{fig:relaxation}). 
The metallicity of (100)-T2, (111)-T3, and -T4 is evident from the band structures shown in Fig.~\ref{fig:bandstructure}: in the slabs cut along the (111) direction, the Fermi energy is crossed by one band, while in the case (100)-T2, two states are partially occupied.
The metallic nature of the (100)-T1 slab is less apparent.
At a glance, the band structure of this system is dominated by a direct gap of about 1.3~eV.
However, a careful inspection reveals that the highest valence band is above the Fermi energy at the $\Gamma$ point and in its vicinity.
Hence, in the 0~K picture provided by DFT, this state is partially unoccupied.
Yet, the energetic proximity with the Fermi energy, on the same order as $k_B T$ at 300~K, suggests that at room temperature the smearing of the Fermi surface may affect the occupation of the bands that are partially unoccupied at 0~K.

\subsubsection{Projected density of states}

\begin{figure*}
	\centering
	\includegraphics[width=0.85\textwidth]{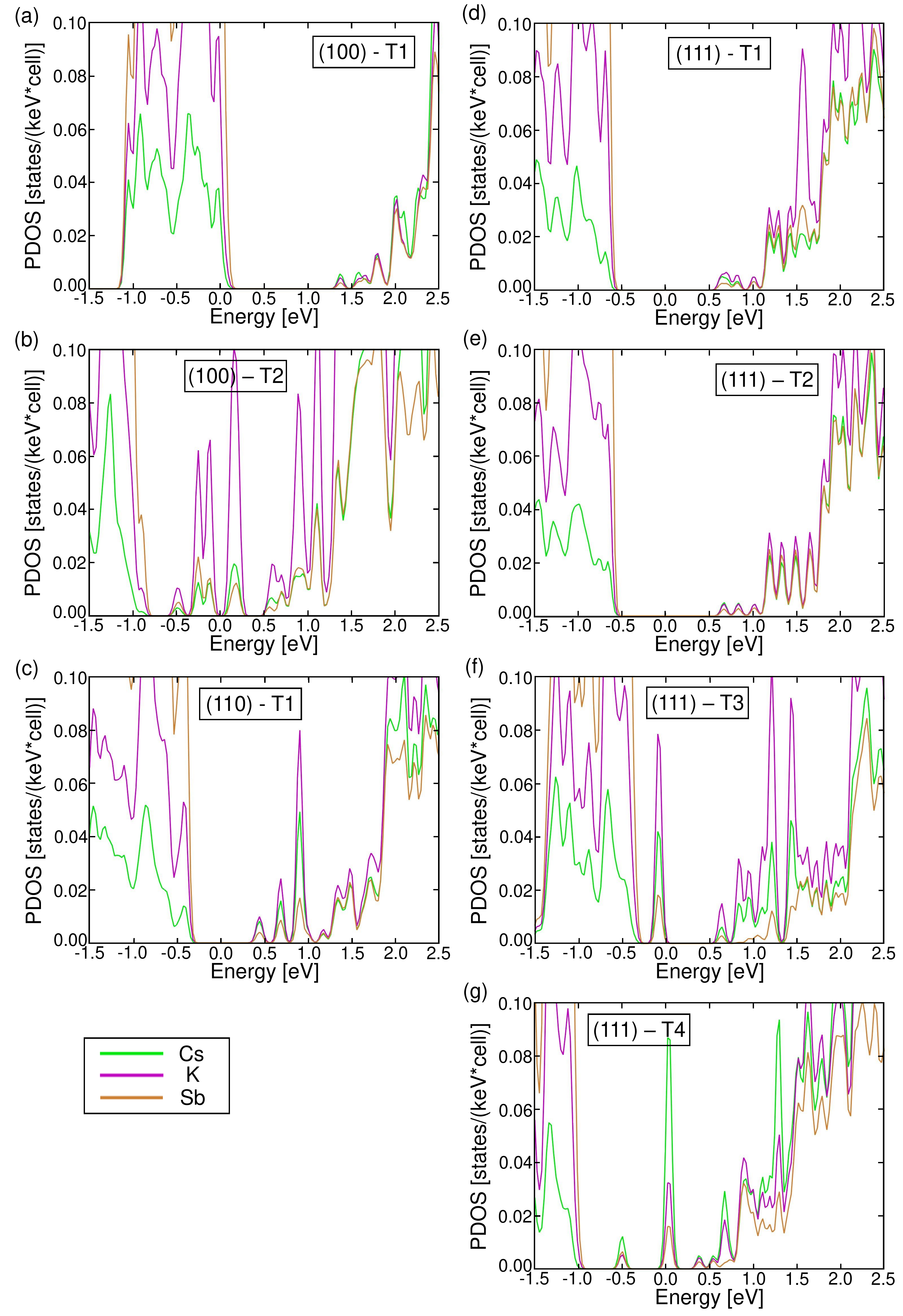}
	\caption{Atom-resolved projected density of states (PDOS) of the considered slabs calculated with the SCAN functional. The energy scale is scaled with respect to the Fermi energy set at 0~eV, which is in the mid-gap of the semiconducting surfaces.}
	\label{fig:pdos}
\end{figure*}

To deepen the analysis of the electronic structure of the considered surfaces, we inspect their atom-resolved PDOS shown in Fig.~\ref{fig:pdos}.
The key characteristics of the bulk~\cite{cocc+18jpcm,cocc+19sr} are largely preserved in all slabs, especially in the semiconducting ones [Fig.~\ref{fig:pdos}a), c), d) and e)]: the top of the valence band is dominated by Sb and K states, while in the conduction region, contributions from the cations prevail. 
In the PDOS of the metallic slabs, the states crossing the Fermi energy exhibit the character of the excess alkali species at the interface with vacuum: K in (100)-T2 and in (111)-T3, Cs in (111)-T4 [Fig.~\ref{fig:pdos}b), f), and g), respectively].
The metallic character of the (100)-T2 surface may not be apparent in Fig.~\ref{fig:pdos}b), where there are no states at the Fermi energy.
However, given the characteristics of the band structure of this system featuring two dispersive states along the M-$\Gamma$-X path (see Fig.~\ref{fig:bandstructure}), and the fact that DFT calculations do not include thermodynamic effects, (100)-T2 should be considered metallic, too.

\subsection{Work Function}

We complete our study with the analysis of the work function of the considered surfaces.
Since both metallic and semiconducting systems are formed, it is important to clarify the definition of this quantity, which is crucial for the assessment of the photoemission performance. 
In metals, the usual definition $W = - E_{Fermi}$ holds straight away: the Fermi energy of the slab, $E_{Fermi}$, obtained as an output of the DFT calculation, is rescaled with respect to the vacuum level.
For the semiconducting surfaces, however, this definition does not strictly apply. In non-metallic systems, the Fermi energy output by DFT calculations is not uniquely defined. Moreover, since the electrons have to overcome the fundamental gap before being ejected from the material, the ionization potential calculated as $IP = - E_{VBM}$ is a more meaningful quantity. $E_{VBM}$ is the energy of the valence band maximum obtained from DFT and offset with respect to the vacuum level.

Examining the results reported in Table~\ref{tab:wf_ip}, we notice some general trends.
With the exception of (111)-T3, metallic surfaces exhibit lower work functions than their semiconducting siblings, in line with physical intuition.
The lowest value overall, $W=2.33$~eV, is found for (100)-T2, which is terminated by a potassium layer, including both inequivalent atoms in the unit cell, K1 and K2, see Fig.~\ref{fig:relaxation}.
Slightly higher values (2.45~eV) are obtained for $IP$ and $W$ (100)-T1 and (111)-T4, respectively: both slabs are terminated by Cs atoms (Fig.~\ref{fig:relaxation}).
The remaining surfaces along the (111) direction, although terminated with K atoms, exhibit lower values of $W$ or $IP$.
While in the (111) slabs with T1 and T2 terminations this behavior can be related to their semiconducting nature, in the case of (111)-T3, we speculate that the substantial atomic rearrangement of the outermost layer plays a crucial role.
Finally, the (110)-T1 slab, which is isostoichiometric with the bulk and which includes all species in the outermost atomic layer, has an intermediate value of $IP=2.83$~eV.
Considering now these results with respect to the surface stability, we notice that all the most favorable systems indicated in Fig.~\ref{fig:energetics} exhibit lower values of $W$ or $IP$ than the remaining ones.

It is meaningful to conclude this analysis by discussing the results reported in Table~\ref{tab:wf_ip} with the available experimental data for the photoemission threshold of CsK$_2$Sb~\cite{ghos-varm78jap,baza+11apl,xie+16prab,feng+17jap}.
All these studies report values close to 2~eV.
Considering the contribution of the absorption process, which is known to lower the emission threshold by a few hundred meV~\cite{spic58pr}, as confirmed by the computed exciton binding energies of bulk CsK$_2$Sb~\cite{cocc+18jpcm}, we can consider our estimates for the most stable surfaces in agreement with experiments. 

Finally, it should be taken into account that empirical values can depart significantly from idealized predictions due to extrinsic conditions, such as surface roughness~\cite{kark-baza15prap} or temperature variations~\cite{mamu+17prab}, which are not included in our DFT calculations.
We speculate that the effect of defects and/or a thermal gradient can modify the picture provided by this analysis but not invalidate it.
As mentioned above, we expect that the smearing of the Fermi surface at room temperature will alter the partial occupation of the valence states in the (100)-T1 slab. 
The presence of defects may affect more significantly the electronic structure, for example, by introducing additional gap states.
To provide a meaningful answer to this question, a dedicated analysis is needed, which, however, goes beyond the scope of this work.

\section{Summary and conclusions}
In summary, we have presented a first-principles work based on DFT on the structural, energetic, and electronic properties of CsK$_2$Sb surfaces with low Miller indices.
Owing to the ternary composition of this material, seven systems with different atomic terminations are obtained cutting the crystal along the directions (100), (110), and (111).
Among them, only the (110) surface naturally retains the same stoichiometry as the parent crystal. 
Upon structural optimization, we found that the outermost atomic layers may relax in a significantly different configuration as the ideal slabs cut off the bulk. 
This occurs in particular for surfaces oriented in the (111) direction, where the electrostatic repulsion generated by the proximity of alkali atoms of both species is minimized by substantial rearrangements. 
From the computed surface energies, we found clear preference for surface formation along the (111) direction, especially at low concentrations of Cs atoms and at simultaneously high concentration of both alkali species.
Under intermediate conditions, the formation of (100) and (110) slabs is favored.
The analysis of the electronic properties of the systems reveals the existence of both metallic and semiconducting surfaces.
Among the latter, we find the most stable slabs, which exhibit band-gap values in excellent agreement with experimental data available for CsK$_2$Sb samples. 
One of the two (100) surfaces is characterized by partially unoccupied valence bands at 0~K, but it is expected that at room temperature the smearing of the Fermi surface will alter this scenario.
Semiconducting slabs have band gaps ranging from 0.8 to about 1.3~eV.
In systems exhibiting an indirect gap, the optical gap is only a few tens of meV larger than its transport counterpart, due to the low dispersion of the highest-occupied bands. 
The analysis of the PDOS reveals similarities between the slabs and the parent bulk regarding the atomic contributions to the electronic states in both valence and conduction regions. 
Moreover, it pinpoints a clear correspondence between the terminating atoms of the metallic slabs and the character of the levels crossing the Fermi energy in these systems.
The work functions (or ionization potentials) calculated for the considered surfaces range from 2.3 to 3.5~eV. 
Interestingly, the lowest values in this range, which are in line with available measurements of photoemission thresholds in CsK$_2$Sb, are found for the most stable surfaces.

In conclusion, the presented results provide an important step forward for the comprehension of the microscopic properties of bi-alkali antimonides.
The gained insight regarding the stability and electronic properties, in particular, is of great relevance for the ongoing activities of growth and characterization of these materials.
For instance, the analysis of the surface stability as a function of the alkali metal concentration is key to understand which type of orientation and termination the sample assumes under specific experimental conditions.
Connecting this knowledge with the electronic structure of the systems is expected to enable efficient strategies to control synthesis in view of targeted applications.
Although idealized, the surfaces modeled in this work offer a realistic glimpse into the characteristics of real samples, especially in terms of atomic structural and electronic properties.
Importantly, the output of these first-principles calculations can be used as an input for the three-step model for photoemission. 

As an outlook, we envision further work to simulate surface morphology in better agreement with experimental samples. 
The inclusion of defects in the structures and of adsorbed gaseous pollutants are just two of the many aspects that have to be addressed in order to provide reliable predictions and characterizations of novel photocathode materials.
Likewise, the inclusion of temperature in the evaluation of the physical properties of the surfaces is essential to closely reproduce the experimental scenario of material growth. 
This parameter is not only crucial for the determination of the photoemission yield~\cite{mamu+17prab,nang+21prb} but it is also expected to influence the stability of the systems and the microscopic characteristics of their electronic structure. 
The results of this work provide an essential starting point to move forward in these directions.

\section*{Acknowledgments}
Stimulating discussions with Thorsten Kamps and Sonal Mistry are gratefully acknowledged. This work was supported by the German Research Foundation (DFG), project number 490940284, by the German Federal Ministry of Education and Research (Professorinnenprogramm III), and by the State of Lower Saxony (Professorinnen für Niedersachsen). Computational resources were provided by the North-German Supercomputing Alliance (HLRN), project nic00069. The authors thank J\"urg Hutter for the permission to use the GTH-pseudopotentials specifically parameterized for the SCAN functional.

\section*{Data Availability}
All data produced in this work, including input and relevant output files, are available free of charge on Zenodo, DOI 10.5281/zenodo.6979440.


\begin{thebibliography}{59}%
\makeatletter
\providecommand \@ifxundefined [1]{%
 \@ifx{#1\undefined}
}%
\providecommand \@ifnum [1]{%
 \ifnum #1\expandafter \@firstoftwo
 \else \expandafter \@secondoftwo
 \fi
}%
\providecommand \@ifx [1]{%
 \ifx #1\expandafter \@firstoftwo
 \else \expandafter \@secondoftwo
 \fi
}%
\providecommand \natexlab [1]{#1}%
\providecommand \enquote  [1]{``#1''}%
\providecommand \bibnamefont  [1]{#1}%
\providecommand \bibfnamefont [1]{#1}%
\providecommand \citenamefont [1]{#1}%
\providecommand \href@noop [0]{\@secondoftwo}%
\providecommand \href [0]{\begingroup \@sanitize@url \@href}%
\providecommand \@href[1]{\@@startlink{#1}\@@href}%
\providecommand \@@href[1]{\endgroup#1\@@endlink}%
\providecommand \@sanitize@url [0]{\catcode `\\12\catcode `\$12\catcode
  `\&12\catcode `\#12\catcode `\^12\catcode `\_12\catcode `\%12\relax}%
\providecommand \@@startlink[1]{}%
\providecommand \@@endlink[0]{}%
\providecommand \url  [0]{\begingroup\@sanitize@url \@url }%
\providecommand \@url [1]{\endgroup\@href {#1}{\urlprefix }}%
\providecommand \urlprefix  [0]{URL }%
\providecommand \Eprint [0]{\href }%
\providecommand \doibase [0]{https://doi.org/}%
\providecommand \selectlanguage [0]{\@gobble}%
\providecommand \bibinfo  [0]{\@secondoftwo}%
\providecommand \bibfield  [0]{\@secondoftwo}%
\providecommand \translation [1]{[#1]}%
\providecommand \BibitemOpen [0]{}%
\providecommand \bibitemStop [0]{}%
\providecommand \bibitemNoStop [0]{.\EOS\space}%
\providecommand \EOS [0]{\spacefactor3000\relax}%
\providecommand \BibitemShut  [1]{\csname bibitem#1\endcsname}%
\let\auto@bib@innerbib\@empty
\bibitem [{\citenamefont {Musumeci}\ \emph {et~al.}(2018)\citenamefont
  {Musumeci}, \citenamefont {Navarro}, \citenamefont {Rosenzweig},
  \citenamefont {Cultrera}, \citenamefont {Bazarov}, \citenamefont {Maxson},
  \citenamefont {Karkare},\ and\ \citenamefont {Padmore}}]{musu+18nimpra}%
  \BibitemOpen
  \bibfield  {author} {\bibinfo {author} {\bibfnamefont {P.}~\bibnamefont
  {Musumeci}}, \bibinfo {author} {\bibfnamefont {J.~G.}\ \bibnamefont
  {Navarro}}, \bibinfo {author} {\bibfnamefont {J.}~\bibnamefont {Rosenzweig}},
  \bibinfo {author} {\bibfnamefont {L.}~\bibnamefont {Cultrera}}, \bibinfo
  {author} {\bibfnamefont {I.}~\bibnamefont {Bazarov}}, \bibinfo {author}
  {\bibfnamefont {J.}~\bibnamefont {Maxson}}, \bibinfo {author} {\bibfnamefont
  {S.}~\bibnamefont {Karkare}},\ and\ \bibinfo {author} {\bibfnamefont
  {H.}~\bibnamefont {Padmore}},\ }\bibfield  {title} {\bibinfo {title}
  {Advances in bright electron sources},\ }\href@noop {} {\bibfield  {journal}
  {\bibinfo  {journal} {Nucl.~Instrum.~Methods~Phys.~Res.~A}\ }\textbf
  {\bibinfo {volume} {907}},\ \bibinfo {pages} {209} (\bibinfo {year}
  {2018})}\BibitemShut {NoStop}%
\bibitem [{\citenamefont {Panuganti}\ \emph {et~al.}(2021)\citenamefont
  {Panuganti}, \citenamefont {Chevallay}, \citenamefont {Fedosseev},\ and\
  \citenamefont {Himmerlich}}]{panu+21nimpra}%
  \BibitemOpen
  \bibfield  {author} {\bibinfo {author} {\bibfnamefont {H.}~\bibnamefont
  {Panuganti}}, \bibinfo {author} {\bibfnamefont {E.}~\bibnamefont
  {Chevallay}}, \bibinfo {author} {\bibfnamefont {V.}~\bibnamefont
  {Fedosseev}},\ and\ \bibinfo {author} {\bibfnamefont {M.}~\bibnamefont
  {Himmerlich}},\ }\bibfield  {title} {\bibinfo {title} {Synthesis, surface
  chemical analysis, lifetime studies and degradation mechanisms of cs-k-sb
  photocathodes},\ }\href@noop {} {\bibfield  {journal} {\bibinfo  {journal}
  {Nucl.~Instrum.~Methods~Phys.~Res.~A}\ }\textbf {\bibinfo {volume} {986}},\
  \bibinfo {pages} {164724} (\bibinfo {year} {2021})}\BibitemShut {NoStop}%
\bibitem [{\citenamefont {Vecchione}\ \emph {et~al.}(2011)\citenamefont
  {Vecchione}, \citenamefont {Ben-Zvi}, \citenamefont {Dowell}, \citenamefont
  {Feng}, \citenamefont {Rao}, \citenamefont {Smedley}, \citenamefont {Wan},\
  and\ \citenamefont {Padmore}}]{vecc+11apl}%
  \BibitemOpen
  \bibfield  {author} {\bibinfo {author} {\bibfnamefont {T.}~\bibnamefont
  {Vecchione}}, \bibinfo {author} {\bibfnamefont {I.}~\bibnamefont {Ben-Zvi}},
  \bibinfo {author} {\bibfnamefont {D.}~\bibnamefont {Dowell}}, \bibinfo
  {author} {\bibfnamefont {J.}~\bibnamefont {Feng}}, \bibinfo {author}
  {\bibfnamefont {T.}~\bibnamefont {Rao}}, \bibinfo {author} {\bibfnamefont
  {J.}~\bibnamefont {Smedley}}, \bibinfo {author} {\bibfnamefont
  {W.}~\bibnamefont {Wan}},\ and\ \bibinfo {author} {\bibfnamefont
  {H.}~\bibnamefont {Padmore}},\ }\bibfield  {title} {\bibinfo {title} {A low
  emittance and high efficiency visible light photocathode for high brightness
  accelerator-based x-ray light sources},\ }\href@noop {} {\bibfield  {journal}
  {\bibinfo  {journal} {Appl.~Phys.~Lett.~}\ }\textbf {\bibinfo {volume}
  {99}},\ \bibinfo {pages} {034103} (\bibinfo {year} {2011})}\BibitemShut
  {NoStop}%
\bibitem [{\citenamefont {Schubert}\ \emph {et~al.}(2013)\citenamefont
  {Schubert}, \citenamefont {Ruiz-Os{\'e}s}, \citenamefont {Ben-Zvi},
  \citenamefont {Kamps}, \citenamefont {Liang}, \citenamefont {Muller},
  \citenamefont {M{\"u}ller}, \citenamefont {Padmore}, \citenamefont {Rao},
  \citenamefont {Tong} \emph {et~al.}}]{schu+13aplm}%
  \BibitemOpen
  \bibfield  {author} {\bibinfo {author} {\bibfnamefont {S.}~\bibnamefont
  {Schubert}}, \bibinfo {author} {\bibfnamefont {M.}~\bibnamefont
  {Ruiz-Os{\'e}s}}, \bibinfo {author} {\bibfnamefont {I.}~\bibnamefont
  {Ben-Zvi}}, \bibinfo {author} {\bibfnamefont {T.}~\bibnamefont {Kamps}},
  \bibinfo {author} {\bibfnamefont {X.}~\bibnamefont {Liang}}, \bibinfo
  {author} {\bibfnamefont {E.}~\bibnamefont {Muller}}, \bibinfo {author}
  {\bibfnamefont {K.}~\bibnamefont {M{\"u}ller}}, \bibinfo {author}
  {\bibfnamefont {H.}~\bibnamefont {Padmore}}, \bibinfo {author} {\bibfnamefont
  {T.}~\bibnamefont {Rao}}, \bibinfo {author} {\bibfnamefont {X.}~\bibnamefont
  {Tong}}, \emph {et~al.},\ }\bibfield  {title} {\bibinfo {title} {Bi-alkali
  antimonide photocathodes for high brightness accelerators},\ }\href@noop {}
  {\bibfield  {journal} {\bibinfo  {journal} {APL~Mater.}\ }\textbf {\bibinfo
  {volume} {1}},\ \bibinfo {pages} {032119} (\bibinfo {year}
  {2013})}\BibitemShut {NoStop}%
\bibitem [{\citenamefont {Schubert}\ \emph {et~al.}(2016)\citenamefont
  {Schubert}, \citenamefont {Wong}, \citenamefont {Feng}, \citenamefont
  {Karkare}, \citenamefont {Padmore}, \citenamefont {Ruiz-Os{\'e}s},
  \citenamefont {Smedley}, \citenamefont {Muller}, \citenamefont {Ding},
  \citenamefont {Gaowei} \emph {et~al.}}]{schu+16jap}%
  \BibitemOpen
  \bibfield  {author} {\bibinfo {author} {\bibfnamefont {S.}~\bibnamefont
  {Schubert}}, \bibinfo {author} {\bibfnamefont {J.}~\bibnamefont {Wong}},
  \bibinfo {author} {\bibfnamefont {J.}~\bibnamefont {Feng}}, \bibinfo {author}
  {\bibfnamefont {S.}~\bibnamefont {Karkare}}, \bibinfo {author} {\bibfnamefont
  {H.}~\bibnamefont {Padmore}}, \bibinfo {author} {\bibfnamefont
  {M.}~\bibnamefont {Ruiz-Os{\'e}s}}, \bibinfo {author} {\bibfnamefont
  {J.}~\bibnamefont {Smedley}}, \bibinfo {author} {\bibfnamefont
  {E.}~\bibnamefont {Muller}}, \bibinfo {author} {\bibfnamefont
  {Z.}~\bibnamefont {Ding}}, \bibinfo {author} {\bibfnamefont {M.}~\bibnamefont
  {Gaowei}}, \emph {et~al.},\ }\bibfield  {title} {\bibinfo {title} {Bi-alkali
  antimonide photocathode growth: An x-ray diffraction study},\ }\href@noop {}
  {\bibfield  {journal} {\bibinfo  {journal} {J.~Appl.~Phys.~}\ }\textbf
  {\bibinfo {volume} {120}},\ \bibinfo {pages} {035303} (\bibinfo {year}
  {2016})}\BibitemShut {NoStop}%
\bibitem [{\citenamefont {Feng}\ \emph {et~al.}(2017)\citenamefont {Feng},
  \citenamefont {Karkare}, \citenamefont {Nasiatka}, \citenamefont {Schubert},
  \citenamefont {Smedley},\ and\ \citenamefont {Padmore}}]{feng+17jap}%
  \BibitemOpen
  \bibfield  {author} {\bibinfo {author} {\bibfnamefont {J.}~\bibnamefont
  {Feng}}, \bibinfo {author} {\bibfnamefont {S.}~\bibnamefont {Karkare}},
  \bibinfo {author} {\bibfnamefont {J.}~\bibnamefont {Nasiatka}}, \bibinfo
  {author} {\bibfnamefont {S.}~\bibnamefont {Schubert}}, \bibinfo {author}
  {\bibfnamefont {J.}~\bibnamefont {Smedley}},\ and\ \bibinfo {author}
  {\bibfnamefont {H.}~\bibnamefont {Padmore}},\ }\bibfield  {title} {\bibinfo
  {title} {Near atomically smooth alkali antimonide photocathode thin films},\
  }\href@noop {} {\bibfield  {journal} {\bibinfo  {journal} {J.~Appl.~Phys.~}\
  }\textbf {\bibinfo {volume} {121}},\ \bibinfo {pages} {044904} (\bibinfo
  {year} {2017})}\BibitemShut {NoStop}%
\bibitem [{\citenamefont {Ding}\ \emph {et~al.}(2017)\citenamefont {Ding},
  \citenamefont {Gaowei}, \citenamefont {Sinsheimer}, \citenamefont {Xie},
  \citenamefont {Schubert}, \citenamefont {Padmore}, \citenamefont {Muller},\
  and\ \citenamefont {Smedley}}]{ding+17jap}%
  \BibitemOpen
  \bibfield  {author} {\bibinfo {author} {\bibfnamefont {Z.}~\bibnamefont
  {Ding}}, \bibinfo {author} {\bibfnamefont {M.}~\bibnamefont {Gaowei}},
  \bibinfo {author} {\bibfnamefont {J.}~\bibnamefont {Sinsheimer}}, \bibinfo
  {author} {\bibfnamefont {J.}~\bibnamefont {Xie}}, \bibinfo {author}
  {\bibfnamefont {S.}~\bibnamefont {Schubert}}, \bibinfo {author}
  {\bibfnamefont {H.}~\bibnamefont {Padmore}}, \bibinfo {author} {\bibfnamefont
  {E.}~\bibnamefont {Muller}},\ and\ \bibinfo {author} {\bibfnamefont
  {J.}~\bibnamefont {Smedley}},\ }\bibfield  {title} {\bibinfo {title} {In-situ
  synchrotron x-ray characterization of k2cssb photocathode grown by ternary
  co-evaporation},\ }\href@noop {} {\bibfield  {journal} {\bibinfo  {journal}
  {J.~Appl.~Phys.~}\ }\textbf {\bibinfo {volume} {121}},\ \bibinfo {pages}
  {055305} (\bibinfo {year} {2017})}\BibitemShut {NoStop}%
\bibitem [{\citenamefont {Gaowei}\ \emph {et~al.}(2017)\citenamefont {Gaowei},
  \citenamefont {Ding}, \citenamefont {Schubert}, \citenamefont {Bhandari},
  \citenamefont {Sinsheimer}, \citenamefont {Kuehn}, \citenamefont {Nagarkar},
  \citenamefont {Marshall}, \citenamefont {Walsh}, \citenamefont {Muller} \emph
  {et~al.}}]{gaow+17aplm}%
  \BibitemOpen
  \bibfield  {author} {\bibinfo {author} {\bibfnamefont {M.}~\bibnamefont
  {Gaowei}}, \bibinfo {author} {\bibfnamefont {Z.}~\bibnamefont {Ding}},
  \bibinfo {author} {\bibfnamefont {S.}~\bibnamefont {Schubert}}, \bibinfo
  {author} {\bibfnamefont {H.}~\bibnamefont {Bhandari}}, \bibinfo {author}
  {\bibfnamefont {J.}~\bibnamefont {Sinsheimer}}, \bibinfo {author}
  {\bibfnamefont {J.}~\bibnamefont {Kuehn}}, \bibinfo {author} {\bibfnamefont
  {V.}~\bibnamefont {Nagarkar}}, \bibinfo {author} {\bibfnamefont
  {M.}~\bibnamefont {Marshall}}, \bibinfo {author} {\bibfnamefont
  {J.}~\bibnamefont {Walsh}}, \bibinfo {author} {\bibfnamefont
  {E.}~\bibnamefont {Muller}}, \emph {et~al.},\ }\bibfield  {title} {\bibinfo
  {title} {Synthesis and x-ray characterization of sputtered bi-alkali
  antimonide photocathodes},\ }\href@noop {} {\bibfield  {journal} {\bibinfo
  {journal} {APL~Mater.}\ }\textbf {\bibinfo {volume} {5}},\ \bibinfo {pages}
  {116104} (\bibinfo {year} {2017})}\BibitemShut {NoStop}%
\bibitem [{\citenamefont {Schmei{\ss}er}\ \emph {et~al.}(2018)\citenamefont
  {Schmei{\ss}er}, \citenamefont {Mistry}, \citenamefont {Kirschner},
  \citenamefont {Schubert}, \citenamefont {Jankowiak}, \citenamefont {Kamps},\
  and\ \citenamefont {K{\"u}hn}}]{schm+18prab}%
  \BibitemOpen
  \bibfield  {author} {\bibinfo {author} {\bibfnamefont {M.~A.}\ \bibnamefont
  {Schmei{\ss}er}}, \bibinfo {author} {\bibfnamefont {S.}~\bibnamefont
  {Mistry}}, \bibinfo {author} {\bibfnamefont {H.}~\bibnamefont {Kirschner}},
  \bibinfo {author} {\bibfnamefont {S.}~\bibnamefont {Schubert}}, \bibinfo
  {author} {\bibfnamefont {A.}~\bibnamefont {Jankowiak}}, \bibinfo {author}
  {\bibfnamefont {T.}~\bibnamefont {Kamps}},\ and\ \bibinfo {author}
  {\bibfnamefont {J.}~\bibnamefont {K{\"u}hn}},\ }\bibfield  {title} {\bibinfo
  {title} {Towards the operation of cs-k-sb photocathodes in superconducting rf
  photoinjectors},\ }\href@noop {} {\bibfield  {journal} {\bibinfo  {journal}
  {Phys. Rev. Accel. Beams}\ }\textbf {\bibinfo {volume} {21}},\ \bibinfo
  {pages} {113401} (\bibinfo {year} {2018})}\BibitemShut {NoStop}%
\bibitem [{\citenamefont {Murtaza}\ \emph {et~al.}(2016)\citenamefont
  {Murtaza}, \citenamefont {Ullah}, \citenamefont {Ullah}, \citenamefont
  {Rani}, \citenamefont {Muzammil}, \citenamefont {Khenata}, \citenamefont
  {Ramay},\ and\ \citenamefont {Khan}}]{murt+16bms}%
  \BibitemOpen
  \bibfield  {author} {\bibinfo {author} {\bibfnamefont {G.}~\bibnamefont
  {Murtaza}}, \bibinfo {author} {\bibfnamefont {M.}~\bibnamefont {Ullah}},
  \bibinfo {author} {\bibfnamefont {N.}~\bibnamefont {Ullah}}, \bibinfo
  {author} {\bibfnamefont {M.}~\bibnamefont {Rani}}, \bibinfo {author}
  {\bibfnamefont {M.}~\bibnamefont {Muzammil}}, \bibinfo {author}
  {\bibfnamefont {R.}~\bibnamefont {Khenata}}, \bibinfo {author} {\bibfnamefont
  {S.~M.}\ \bibnamefont {Ramay}},\ and\ \bibinfo {author} {\bibfnamefont
  {U.}~\bibnamefont {Khan}},\ }\bibfield  {title} {\bibinfo {title}
  {Structural, elastic, electronic and optical properties of bi-alkali
  antimonides},\ }\href@noop {} {\bibfield  {journal} {\bibinfo  {journal}
  {Bull.~Mater.~Sci.}\ }\textbf {\bibinfo {volume} {39}},\ \bibinfo {pages}
  {1581} (\bibinfo {year} {2016})}\BibitemShut {NoStop}%
\bibitem [{\citenamefont {Kalarasse}\ \emph
  {et~al.}(2010{\natexlab{a}})\citenamefont {Kalarasse}, \citenamefont
  {Bennecer},\ and\ \citenamefont {Kalarasse}}]{kala+10jpcs}%
  \BibitemOpen
  \bibfield  {author} {\bibinfo {author} {\bibfnamefont {L.}~\bibnamefont
  {Kalarasse}}, \bibinfo {author} {\bibfnamefont {B.}~\bibnamefont
  {Bennecer}},\ and\ \bibinfo {author} {\bibfnamefont {F.}~\bibnamefont
  {Kalarasse}},\ }\bibfield  {title} {\bibinfo {title} {Optical properties of
  the alkali antimonide semiconductors cs3sb, cs2ksb, csk2sb and k3sb},\
  }\href@noop {} {\bibfield  {journal} {\bibinfo  {journal}
  {J.~Phys.~Chem.~Solids}\ }\textbf {\bibinfo {volume} {71}},\ \bibinfo {pages}
  {314} (\bibinfo {year} {2010}{\natexlab{a}})}\BibitemShut {NoStop}%
\bibitem [{\citenamefont {Kalarasse}\ \emph
  {et~al.}(2010{\natexlab{b}})\citenamefont {Kalarasse}, \citenamefont
  {Bennecer}, \citenamefont {Kalarasse},\ and\ \citenamefont
  {Djeroud}}]{kala+10jpcs1}%
  \BibitemOpen
  \bibfield  {author} {\bibinfo {author} {\bibfnamefont {L.}~\bibnamefont
  {Kalarasse}}, \bibinfo {author} {\bibfnamefont {B.}~\bibnamefont {Bennecer}},
  \bibinfo {author} {\bibfnamefont {F.}~\bibnamefont {Kalarasse}},\ and\
  \bibinfo {author} {\bibfnamefont {S.}~\bibnamefont {Djeroud}},\ }\bibfield
  {title} {\bibinfo {title} {Pressure effect on the electronic and optical
  properties of the alkali antimonide semiconductors cs3sb, kcs2sb, csk2sb and
  k3sb: Ab initio study},\ }\href@noop {} {\bibfield  {journal} {\bibinfo
  {journal} {J.~Phys.~Chem.~Solids}\ }\textbf {\bibinfo {volume} {71}},\
  \bibinfo {pages} {1732} (\bibinfo {year} {2010}{\natexlab{b}})}\BibitemShut
  {NoStop}%
\bibitem [{\citenamefont {Cocchi}\ \emph {et~al.}(2018)\citenamefont {Cocchi},
  \citenamefont {Mistry}, \citenamefont {Schmei{\ss}er}, \citenamefont
  {K{\"u}hn},\ and\ \citenamefont {Kamps}}]{cocc+18jpcm}%
  \BibitemOpen
  \bibfield  {author} {\bibinfo {author} {\bibfnamefont {C.}~\bibnamefont
  {Cocchi}}, \bibinfo {author} {\bibfnamefont {S.}~\bibnamefont {Mistry}},
  \bibinfo {author} {\bibfnamefont {M.}~\bibnamefont {Schmei{\ss}er}}, \bibinfo
  {author} {\bibfnamefont {J.}~\bibnamefont {K{\"u}hn}},\ and\ \bibinfo
  {author} {\bibfnamefont {T.}~\bibnamefont {Kamps}},\ }\bibfield  {title}
  {\bibinfo {title} {First-principles many-body study of the electronic and
  optical properties of csk2sb, a semiconducting material for ultra-bright
  electron sources},\ }\href@noop {} {\bibfield  {journal} {\bibinfo  {journal}
  {J.~Phys.:~Condens.~Matter.~}\ }\textbf {\bibinfo {volume} {31}},\ \bibinfo
  {pages} {014002} (\bibinfo {year} {2018})}\BibitemShut {NoStop}%
\bibitem [{\citenamefont {Cocchi}\ \emph {et~al.}(2019)\citenamefont {Cocchi},
  \citenamefont {Mistry}, \citenamefont {Schmei{\ss}er}, \citenamefont
  {Amador}, \citenamefont {K{\"u}hn},\ and\ \citenamefont {Kamps}}]{cocc+19sr}%
  \BibitemOpen
  \bibfield  {author} {\bibinfo {author} {\bibfnamefont {C.}~\bibnamefont
  {Cocchi}}, \bibinfo {author} {\bibfnamefont {S.}~\bibnamefont {Mistry}},
  \bibinfo {author} {\bibfnamefont {M.}~\bibnamefont {Schmei{\ss}er}}, \bibinfo
  {author} {\bibfnamefont {R.}~\bibnamefont {Amador}}, \bibinfo {author}
  {\bibfnamefont {J.}~\bibnamefont {K{\"u}hn}},\ and\ \bibinfo {author}
  {\bibfnamefont {T.}~\bibnamefont {Kamps}},\ }\bibfield  {title} {\bibinfo
  {title} {Electronic structure and core electron fingerprints of caesium-based
  multi-alkali antimonides for ultra-bright electron sources},\ }\href@noop {}
  {\bibfield  {journal} {\bibinfo  {journal} {Sci.~Rep.~}\ }\textbf {\bibinfo
  {volume} {9}},\ \bibinfo {pages} {1} (\bibinfo {year} {2019})}\BibitemShut
  {NoStop}%
\bibitem [{\citenamefont {Cocchi}(2020)}]{cocc20pssrrl}%
  \BibitemOpen
  \bibfield  {author} {\bibinfo {author} {\bibfnamefont {C.}~\bibnamefont
  {Cocchi}},\ }\bibfield  {title} {\bibinfo {title} {X-ray absorption
  fingerprints from cs atoms in cs3sb},\ }\href
  {https://doi.org/10.1002/pssr.202000194} {\bibfield  {journal} {\bibinfo
  {journal} {Phys.~Status~Solidi~(RRL)}\ }\textbf {\bibinfo {volume} {14}},\
  \bibinfo {pages} {2000194} (\bibinfo {year} {2020})}\BibitemShut {NoStop}%
\bibitem [{\citenamefont {Amador}\ \emph {et~al.}(2021)\citenamefont {Amador},
  \citenamefont {Sa{\ss}nick},\ and\ \citenamefont {Cocchi}}]{amad+21jpcm}%
  \BibitemOpen
  \bibfield  {author} {\bibinfo {author} {\bibfnamefont {R.}~\bibnamefont
  {Amador}}, \bibinfo {author} {\bibfnamefont {H.-D.}\ \bibnamefont
  {Sa{\ss}nick}},\ and\ \bibinfo {author} {\bibfnamefont {C.}~\bibnamefont
  {Cocchi}},\ }\bibfield  {title} {\bibinfo {title} {Electronic structure and
  optical properties of na$_2$ksb and nak$_2$sb from first-principles many-body
  theory},\ }\href {https://doi.org/10.1088/1361-648x/ac0e70} {\bibfield
  {journal} {\bibinfo  {journal} {J.~Phys.:~Condens.~Matter.~}\ }\textbf
  {\bibinfo {volume} {33}},\ \bibinfo {pages} {365502} (\bibinfo {year}
  {2021})}\BibitemShut {NoStop}%
\bibitem [{\citenamefont {Cocchi}\ and\ \citenamefont
  {Sa{\ss}nick}(2021)}]{cocc-sass21micromachines}%
  \BibitemOpen
  \bibfield  {author} {\bibinfo {author} {\bibfnamefont {C.}~\bibnamefont
  {Cocchi}}\ and\ \bibinfo {author} {\bibfnamefont {H.-D.}\ \bibnamefont
  {Sa{\ss}nick}},\ }\bibfield  {title} {\bibinfo {title} {Ab initio
  quantum-mechanical predictions of semiconducting photocathode materials},\
  }\href@noop {} {\bibfield  {journal} {\bibinfo  {journal} {Micromachines}\
  }\textbf {\bibinfo {volume} {12}},\ \bibinfo {pages} {1002} (\bibinfo {year}
  {2021})}\BibitemShut {NoStop}%
\bibitem [{\citenamefont {Khan}\ \emph {et~al.}(2021)\citenamefont {Khan},
  \citenamefont {Murtaza}, \citenamefont {Khan}, \citenamefont {Laref},
  \citenamefont {Kattan},\ and\ \citenamefont {Haneef}}]{khan+21ijer}%
  \BibitemOpen
  \bibfield  {author} {\bibinfo {author} {\bibfnamefont {Z.}~\bibnamefont
  {Khan}}, \bibinfo {author} {\bibfnamefont {G.}~\bibnamefont {Murtaza}},
  \bibinfo {author} {\bibfnamefont {A.~A.}\ \bibnamefont {Khan}}, \bibinfo
  {author} {\bibfnamefont {A.}~\bibnamefont {Laref}}, \bibinfo {author}
  {\bibfnamefont {N.~A.}\ \bibnamefont {Kattan}},\ and\ \bibinfo {author}
  {\bibfnamefont {M.}~\bibnamefont {Haneef}},\ }\bibfield  {title} {\bibinfo
  {title} {Different physical properties of bi-alkali pnictogen compounds using
  density functional theory},\ }\href@noop {} {\bibfield  {journal} {\bibinfo
  {journal} {Int.~J.~Energy~Res.}\ }\textbf {\bibinfo {volume} {45}},\ \bibinfo
  {pages} {7703} (\bibinfo {year} {2021})}\BibitemShut {NoStop}%
\bibitem [{\citenamefont {Sa{\ss}nick}\ and\ \citenamefont
  {Cocchi}(2021)}]{sass-cocc21es}%
  \BibitemOpen
  \bibfield  {author} {\bibinfo {author} {\bibfnamefont {H.-D.}\ \bibnamefont
  {Sa{\ss}nick}}\ and\ \bibinfo {author} {\bibfnamefont {C.}~\bibnamefont
  {Cocchi}},\ }\bibfield  {title} {\bibinfo {title} {Electronic structure of
  cesium-based photocathode materials from density functional theory:
  performance of {PBE}, {SCAN}, and {HSE}06 functionals},\ }\href
  {https://doi.org/10.1088/2516-1075/abfb08} {\bibfield  {journal} {\bibinfo
  {journal} {Electr.~Struct.}\ }\textbf {\bibinfo {volume} {3}},\ \bibinfo
  {pages} {027001} (\bibinfo {year} {2021})}\BibitemShut {NoStop}%
\bibitem [{\citenamefont {Shu}\ \emph {et~al.}(2021)\citenamefont {Shu},
  \citenamefont {Zhang}, \citenamefont {Ren}, \citenamefont {Wang},
  \citenamefont {Jin}, \citenamefont {Zhang},\ and\ \citenamefont
  {Zhan}}]{shu+21optik}%
  \BibitemOpen
  \bibfield  {author} {\bibinfo {author} {\bibfnamefont {Z.}~\bibnamefont
  {Shu}}, \bibinfo {author} {\bibfnamefont {Y.}~\bibnamefont {Zhang}}, \bibinfo
  {author} {\bibfnamefont {L.}~\bibnamefont {Ren}}, \bibinfo {author}
  {\bibfnamefont {X.}~\bibnamefont {Wang}}, \bibinfo {author} {\bibfnamefont
  {M.}~\bibnamefont {Jin}}, \bibinfo {author} {\bibfnamefont {K.}~\bibnamefont
  {Zhang}},\ and\ \bibinfo {author} {\bibfnamefont {J.}~\bibnamefont {Zhan}},\
  }\bibfield  {title} {\bibinfo {title} {Effect of vacancy defects on
  photoelectric properties of k$_2$cssb photocathode},\ }\href@noop {}
  {\bibfield  {journal} {\bibinfo  {journal} {Optik}\ }\textbf {\bibinfo
  {volume} {232}},\ \bibinfo {pages} {166555} (\bibinfo {year}
  {2021})}\BibitemShut {NoStop}%
\bibitem [{\citenamefont {Yamaguchi}\ \emph {et~al.}(2018)\citenamefont
  {Yamaguchi}, \citenamefont {Liu}, \citenamefont {DeFazio}, \citenamefont
  {Gaowei}, \citenamefont {Narvaez~Villarrubia}, \citenamefont {Xie},
  \citenamefont {Sinsheimer}, \citenamefont {Strom}, \citenamefont {Pavlenko},
  \citenamefont {Jensen} \emph {et~al.}}]{yama+18ami}%
  \BibitemOpen
  \bibfield  {author} {\bibinfo {author} {\bibfnamefont {H.}~\bibnamefont
  {Yamaguchi}}, \bibinfo {author} {\bibfnamefont {F.}~\bibnamefont {Liu}},
  \bibinfo {author} {\bibfnamefont {J.}~\bibnamefont {DeFazio}}, \bibinfo
  {author} {\bibfnamefont {M.}~\bibnamefont {Gaowei}}, \bibinfo {author}
  {\bibfnamefont {C.~W.}\ \bibnamefont {Narvaez~Villarrubia}}, \bibinfo
  {author} {\bibfnamefont {J.}~\bibnamefont {Xie}}, \bibinfo {author}
  {\bibfnamefont {J.}~\bibnamefont {Sinsheimer}}, \bibinfo {author}
  {\bibfnamefont {D.}~\bibnamefont {Strom}}, \bibinfo {author} {\bibfnamefont
  {V.}~\bibnamefont {Pavlenko}}, \bibinfo {author} {\bibfnamefont {K.~L.}\
  \bibnamefont {Jensen}}, \emph {et~al.},\ }\bibfield  {title} {\bibinfo
  {title} {Free-standing bialkali photocathodes using atomically thin
  substrates},\ }\href@noop {} {\bibfield  {journal} {\bibinfo  {journal}
  {Adv.~Mater.~Interfaces}\ }\textbf {\bibinfo {volume} {5}},\ \bibinfo {pages}
  {1800249} (\bibinfo {year} {2018})}\BibitemShut {NoStop}%
\bibitem [{\citenamefont {Galdi}\ \emph {et~al.}(2019)\citenamefont {Galdi},
  \citenamefont {Balajka}, \citenamefont {Baretz}, \citenamefont {Bazarov},
  \citenamefont {Cultrera}, \citenamefont {DeBenedetti}, \citenamefont {Hines},
  \citenamefont {Ikponwmen}, \citenamefont {Maxson},\ and\ \citenamefont
  {McBride}}]{gald+19quantum}%
  \BibitemOpen
  \bibfield  {author} {\bibinfo {author} {\bibfnamefont {A.}~\bibnamefont
  {Galdi}}, \bibinfo {author} {\bibfnamefont {J.}~\bibnamefont {Balajka}},
  \bibinfo {author} {\bibfnamefont {J.}~\bibnamefont {Baretz}}, \bibinfo
  {author} {\bibfnamefont {I.}~\bibnamefont {Bazarov}}, \bibinfo {author}
  {\bibfnamefont {L.}~\bibnamefont {Cultrera}}, \bibinfo {author}
  {\bibfnamefont {W.}~\bibnamefont {DeBenedetti}}, \bibinfo {author}
  {\bibfnamefont {M.}~\bibnamefont {Hines}}, \bibinfo {author} {\bibfnamefont
  {F.}~\bibnamefont {Ikponwmen}}, \bibinfo {author} {\bibfnamefont
  {J.}~\bibnamefont {Maxson}},\ and\ \bibinfo {author} {\bibfnamefont
  {S.}~\bibnamefont {McBride}},\ }\bibfield  {title} {\bibinfo {title} {Towards
  the optimization of photocathode properties via surface science techniques: A
  study on cs$_3$sb thin film growth},\ }\href@noop {} {\bibfield  {journal}
  {\bibinfo  {journal} {Quantum}\ }\textbf {\bibinfo {volume} {10}},\ \bibinfo
  {pages} {10} (\bibinfo {year} {2019})}\BibitemShut {NoStop}%
\bibitem [{\citenamefont {Xie}\ \emph {et~al.}(2017)\citenamefont {Xie},
  \citenamefont {Demarteau}, \citenamefont {Wagner}, \citenamefont {Schubert},
  \citenamefont {Gaowei}, \citenamefont {Attenkofer}, \citenamefont {Walsh},
  \citenamefont {Smedley}, \citenamefont {Wong}, \citenamefont {Feng} \emph
  {et~al.}}]{xie+17jpd}%
  \BibitemOpen
  \bibfield  {author} {\bibinfo {author} {\bibfnamefont {J.}~\bibnamefont
  {Xie}}, \bibinfo {author} {\bibfnamefont {M.}~\bibnamefont {Demarteau}},
  \bibinfo {author} {\bibfnamefont {R.}~\bibnamefont {Wagner}}, \bibinfo
  {author} {\bibfnamefont {S.}~\bibnamefont {Schubert}}, \bibinfo {author}
  {\bibfnamefont {M.}~\bibnamefont {Gaowei}}, \bibinfo {author} {\bibfnamefont
  {K.}~\bibnamefont {Attenkofer}}, \bibinfo {author} {\bibfnamefont
  {J.}~\bibnamefont {Walsh}}, \bibinfo {author} {\bibfnamefont
  {J.}~\bibnamefont {Smedley}}, \bibinfo {author} {\bibfnamefont
  {J.}~\bibnamefont {Wong}}, \bibinfo {author} {\bibfnamefont {J.}~\bibnamefont
  {Feng}}, \emph {et~al.},\ }\bibfield  {title} {\bibinfo {title} {Synchrotron
  x-ray study of a low roughness and high efficiency k$_2$cssb photocathode
  during film growth},\ }\href@noop {} {\bibfield  {journal} {\bibinfo
  {journal} {J.~Phys.~D}\ }\textbf {\bibinfo {volume} {50}},\ \bibinfo {pages}
  {205303} (\bibinfo {year} {2017})}\BibitemShut {NoStop}%
\bibitem [{\citenamefont {Gevorkyan}\ \emph {et~al.}(2018)\citenamefont
  {Gevorkyan}, \citenamefont {Karkare}, \citenamefont {Emamian}, \citenamefont
  {Bazarov},\ and\ \citenamefont {Padmore}}]{gevo+18prab}%
  \BibitemOpen
  \bibfield  {author} {\bibinfo {author} {\bibfnamefont {G.}~\bibnamefont
  {Gevorkyan}}, \bibinfo {author} {\bibfnamefont {S.}~\bibnamefont {Karkare}},
  \bibinfo {author} {\bibfnamefont {S.}~\bibnamefont {Emamian}}, \bibinfo
  {author} {\bibfnamefont {I.}~\bibnamefont {Bazarov}},\ and\ \bibinfo {author}
  {\bibfnamefont {H.}~\bibnamefont {Padmore}},\ }\bibfield  {title} {\bibinfo
  {title} {Effects of physical and chemical surface roughness on the brightness
  of electron beams from photocathodes},\ }\href@noop {} {\bibfield  {journal}
  {\bibinfo  {journal} {Phys. Rev. Accel. Beams}\ }\textbf {\bibinfo {volume}
  {21}},\ \bibinfo {pages} {093401} (\bibinfo {year} {2018})}\BibitemShut
  {NoStop}%
\bibitem [{\citenamefont {Galdi}\ \emph {et~al.}(2021)\citenamefont {Galdi},
  \citenamefont {Balajka}, \citenamefont {DeBenedetti}, \citenamefont
  {Cultrera}, \citenamefont {Bazarov}, \citenamefont {Hines},\ and\
  \citenamefont {Maxson}}]{gald+21apl}%
  \BibitemOpen
  \bibfield  {author} {\bibinfo {author} {\bibfnamefont {A.}~\bibnamefont
  {Galdi}}, \bibinfo {author} {\bibfnamefont {J.}~\bibnamefont {Balajka}},
  \bibinfo {author} {\bibfnamefont {W.~J.~I.}\ \bibnamefont {DeBenedetti}},
  \bibinfo {author} {\bibfnamefont {L.}~\bibnamefont {Cultrera}}, \bibinfo
  {author} {\bibfnamefont {I.~V.}\ \bibnamefont {Bazarov}}, \bibinfo {author}
  {\bibfnamefont {M.~A.}\ \bibnamefont {Hines}},\ and\ \bibinfo {author}
  {\bibfnamefont {J.~M.}\ \bibnamefont {Maxson}},\ }\bibfield  {title}
  {\bibinfo {title} {Reduction of surface roughness emittance of cs3sb
  photocathodes grown via codeposition on single crystal substrates},\ }\href
  {https://doi.org/10.1063/5.0053186} {\bibfield  {journal} {\bibinfo
  {journal} {Appl.~Phys.~Lett.~}\ }\textbf {\bibinfo {volume} {118}},\ \bibinfo
  {pages} {244101} (\bibinfo {year} {2021})},\ \Eprint
  {https://arxiv.org/abs/https://doi.org/10.1063/5.0053186}
  {https://doi.org/10.1063/5.0053186} \BibitemShut {NoStop}%
\bibitem [{\citenamefont {Berglund}\ and\ \citenamefont
  {Spicer}(1964)}]{berg-spic64pr}%
  \BibitemOpen
  \bibfield  {author} {\bibinfo {author} {\bibfnamefont {C.~N.}\ \bibnamefont
  {Berglund}}\ and\ \bibinfo {author} {\bibfnamefont {W.~E.}\ \bibnamefont
  {Spicer}},\ }\bibfield  {title} {\bibinfo {title} {Photoemission studies of
  copper and silver: theory},\ }\href@noop {} {\bibfield  {journal} {\bibinfo
  {journal} {Phys.~Rev.~}\ }\textbf {\bibinfo {volume} {136}},\ \bibinfo
  {pages} {A1030} (\bibinfo {year} {1964})}\BibitemShut {NoStop}%
\bibitem [{\citenamefont {Antoniuk}\ \emph {et~al.}(2020)\citenamefont
  {Antoniuk}, \citenamefont {Yue}, \citenamefont {Zhou}, \citenamefont
  {Schindler}, \citenamefont {Schroeder}, \citenamefont {Dunham}, \citenamefont
  {Pianetta}, \citenamefont {Vecchione},\ and\ \citenamefont
  {Reed}}]{anto+20prb}%
  \BibitemOpen
  \bibfield  {author} {\bibinfo {author} {\bibfnamefont {E.~R.}\ \bibnamefont
  {Antoniuk}}, \bibinfo {author} {\bibfnamefont {Y.}~\bibnamefont {Yue}},
  \bibinfo {author} {\bibfnamefont {Y.}~\bibnamefont {Zhou}}, \bibinfo {author}
  {\bibfnamefont {P.}~\bibnamefont {Schindler}}, \bibinfo {author}
  {\bibfnamefont {W.~A.}\ \bibnamefont {Schroeder}}, \bibinfo {author}
  {\bibfnamefont {B.}~\bibnamefont {Dunham}}, \bibinfo {author} {\bibfnamefont
  {P.}~\bibnamefont {Pianetta}}, \bibinfo {author} {\bibfnamefont
  {T.}~\bibnamefont {Vecchione}},\ and\ \bibinfo {author} {\bibfnamefont
  {E.~J.}\ \bibnamefont {Reed}},\ }\bibfield  {title} {\bibinfo {title}
  {Generalizable density functional theory based photoemission model for the
  accelerated development of photocathodes and other photoemissive devices},\
  }\href {https://doi.org/10.1103/PhysRevB.101.235447} {\bibfield  {journal}
  {\bibinfo  {journal} {Phys.~Rev.~B}\ }\textbf {\bibinfo {volume} {101}},\
  \bibinfo {pages} {235447} (\bibinfo {year} {2020})}\BibitemShut {NoStop}%
\bibitem [{\citenamefont {Nangoi}\ \emph {et~al.}(2021)\citenamefont {Nangoi},
  \citenamefont {Karkare}, \citenamefont {Sundararaman}, \citenamefont
  {Padmore},\ and\ \citenamefont {Arias}}]{nang+21prb}%
  \BibitemOpen
  \bibfield  {author} {\bibinfo {author} {\bibfnamefont {J.~K.}\ \bibnamefont
  {Nangoi}}, \bibinfo {author} {\bibfnamefont {S.}~\bibnamefont {Karkare}},
  \bibinfo {author} {\bibfnamefont {R.}~\bibnamefont {Sundararaman}}, \bibinfo
  {author} {\bibfnamefont {H.~A.}\ \bibnamefont {Padmore}},\ and\ \bibinfo
  {author} {\bibfnamefont {T.~A.}\ \bibnamefont {Arias}},\ }\bibfield  {title}
  {\bibinfo {title} {Importance of bulk excitations and coherent
  electron-photon-phonon scattering in photoemission from pbte(111): Ab initio
  theory with experimental comparisons},\ }\href
  {https://doi.org/10.1103/PhysRevB.104.115132} {\bibfield  {journal} {\bibinfo
   {journal} {Phys.~Rev.~B}\ }\textbf {\bibinfo {volume} {104}},\ \bibinfo
  {pages} {115132} (\bibinfo {year} {2021})}\BibitemShut {NoStop}%
\bibitem [{\citenamefont {Mamun}\ \emph {et~al.}(2017)\citenamefont {Mamun},
  \citenamefont {Hernandez-Flores}, \citenamefont {Morales}, \citenamefont
  {Hernandez-Garcia},\ and\ \citenamefont {Poelker}}]{mamu+17prab}%
  \BibitemOpen
  \bibfield  {author} {\bibinfo {author} {\bibfnamefont {M.}~\bibnamefont
  {Mamun}}, \bibinfo {author} {\bibfnamefont {M.}~\bibnamefont
  {Hernandez-Flores}}, \bibinfo {author} {\bibfnamefont {E.}~\bibnamefont
  {Morales}}, \bibinfo {author} {\bibfnamefont {C.}~\bibnamefont
  {Hernandez-Garcia}},\ and\ \bibinfo {author} {\bibfnamefont {M.}~\bibnamefont
  {Poelker}},\ }\bibfield  {title} {\bibinfo {title} {Temperature dependence of
  alkali-antimonide photocathodes: Evaluation at cryogenic temperatures},\
  }\href@noop {} {\bibfield  {journal} {\bibinfo  {journal} {Phys. Rev. Accel.
  Beams}\ }\textbf {\bibinfo {volume} {20}},\ \bibinfo {pages} {103403}
  (\bibinfo {year} {2017})}\BibitemShut {NoStop}%
\bibitem [{\citenamefont {Dai}\ \emph {et~al.}(2020)\citenamefont {Dai},
  \citenamefont {Ding}, \citenamefont {Ruan}, \citenamefont {Xu},\ and\
  \citenamefont {Liu}}]{dai+20}%
  \BibitemOpen
  \bibfield  {author} {\bibinfo {author} {\bibfnamefont {J.}~\bibnamefont
  {Dai}}, \bibinfo {author} {\bibfnamefont {Y.}~\bibnamefont {Ding}}, \bibinfo
  {author} {\bibfnamefont {C.}~\bibnamefont {Ruan}}, \bibinfo {author}
  {\bibfnamefont {X.}~\bibnamefont {Xu}},\ and\ \bibinfo {author}
  {\bibfnamefont {H.}~\bibnamefont {Liu}},\ }\bibfield  {title} {\bibinfo
  {title} {High photocurrent density and continuous electron emission
  characterization of a multi-alkali antimonide photocathode},\ }\href@noop {}
  {\bibfield  {journal} {\bibinfo  {journal} {Electronics}\ }\textbf {\bibinfo
  {volume} {9}},\ \bibinfo {pages} {1991} (\bibinfo {year} {2020})}\BibitemShut
  {NoStop}%
\bibitem [{\citenamefont {Parzyck}\ \emph {et~al.}(2022)\citenamefont
  {Parzyck}, \citenamefont {Galdi}, \citenamefont {Nangoi}, \citenamefont
  {DeBenedetti}, \citenamefont {Balajka}, \citenamefont {Faeth}, \citenamefont
  {Paik}, \citenamefont {Hu}, \citenamefont {Arias}, \citenamefont {Hines}
  \emph {et~al.}}]{parz+22prl}%
  \BibitemOpen
  \bibfield  {author} {\bibinfo {author} {\bibfnamefont {C.~T.}\ \bibnamefont
  {Parzyck}}, \bibinfo {author} {\bibfnamefont {A.}~\bibnamefont {Galdi}},
  \bibinfo {author} {\bibfnamefont {J.~K.}\ \bibnamefont {Nangoi}}, \bibinfo
  {author} {\bibfnamefont {W.~J.~I.}\ \bibnamefont {DeBenedetti}}, \bibinfo
  {author} {\bibfnamefont {J.}~\bibnamefont {Balajka}}, \bibinfo {author}
  {\bibfnamefont {B.~D.}\ \bibnamefont {Faeth}}, \bibinfo {author}
  {\bibfnamefont {H.}~\bibnamefont {Paik}}, \bibinfo {author} {\bibfnamefont
  {C.}~\bibnamefont {Hu}}, \bibinfo {author} {\bibfnamefont {T.~A.}\
  \bibnamefont {Arias}}, \bibinfo {author} {\bibfnamefont {M.~A.}\ \bibnamefont
  {Hines}}, \emph {et~al.},\ }\bibfield  {title} {\bibinfo {title}
  {Single-crystal alkali antimonide photocathodes: High efficiency in the
  ultrathin limit},\ }\href {https://doi.org/10.1103/PhysRevLett.128.114801}
  {\bibfield  {journal} {\bibinfo  {journal} {Phys.~Rev.~Lett.~}\ }\textbf
  {\bibinfo {volume} {128}},\ \bibinfo {pages} {114801} (\bibinfo {year}
  {2022})}\BibitemShut {NoStop}%
\bibitem [{\citenamefont {Hohenberg}\ and\ \citenamefont
  {Kohn}(1964)}]{hohe-kohn64pr}%
  \BibitemOpen
  \bibfield  {author} {\bibinfo {author} {\bibfnamefont {P.}~\bibnamefont
  {Hohenberg}}\ and\ \bibinfo {author} {\bibfnamefont {W.}~\bibnamefont
  {Kohn}},\ }\bibfield  {title} {\bibinfo {title} {Inhomogeneus electron gas},\
  }\href@noop {} {\bibfield  {journal} {\bibinfo  {journal} {Phys.~Rev.~}\
  }\textbf {\bibinfo {volume} {136}},\ \bibinfo {pages} {B864} (\bibinfo {year}
  {1964})}\BibitemShut {NoStop}%
\bibitem [{\citenamefont {Kohn}\ and\ \citenamefont
  {Sham}(1965)}]{kohn-sham65pr}%
  \BibitemOpen
  \bibfield  {author} {\bibinfo {author} {\bibfnamefont {W.}~\bibnamefont
  {Kohn}}\ and\ \bibinfo {author} {\bibfnamefont {L.~J.}\ \bibnamefont
  {Sham}},\ }\bibfield  {title} {\bibinfo {title} {Self-consistent equations
  including exchange and correlation effects},\ }\href@noop {} {\bibfield
  {journal} {\bibinfo  {journal} {Phys.~Rev.~}\ }\textbf {\bibinfo {volume}
  {140}},\ \bibinfo {pages} {A1133} (\bibinfo {year} {1965})}\BibitemShut
  {NoStop}%
\bibitem [{\citenamefont {Kühne}\ \emph {et~al.}(2020)\citenamefont {Kühne},
  \citenamefont {Iannuzzi}, \citenamefont {Del~Ben}, \citenamefont {Rybkin},
  \citenamefont {Seewald}, \citenamefont {Stein}, \citenamefont {Laino},
  \citenamefont {Khaliullin}, \citenamefont {Schütt}, \citenamefont
  {Schiffmann} \emph {et~al.}}]{kueh+20jcp}%
  \BibitemOpen
  \bibfield  {author} {\bibinfo {author} {\bibfnamefont {T.~D.}\ \bibnamefont
  {Kühne}}, \bibinfo {author} {\bibfnamefont {M.}~\bibnamefont {Iannuzzi}},
  \bibinfo {author} {\bibfnamefont {M.}~\bibnamefont {Del~Ben}}, \bibinfo
  {author} {\bibfnamefont {V.~V.}\ \bibnamefont {Rybkin}}, \bibinfo {author}
  {\bibfnamefont {P.}~\bibnamefont {Seewald}}, \bibinfo {author} {\bibfnamefont
  {F.}~\bibnamefont {Stein}}, \bibinfo {author} {\bibfnamefont
  {T.}~\bibnamefont {Laino}}, \bibinfo {author} {\bibfnamefont {R.~Z.}\
  \bibnamefont {Khaliullin}}, \bibinfo {author} {\bibfnamefont
  {O.}~\bibnamefont {Schütt}}, \bibinfo {author} {\bibfnamefont
  {F.}~\bibnamefont {Schiffmann}}, \emph {et~al.},\ }\bibfield  {title}
  {\bibinfo {title} {Cp2k: An electronic structure and molecular dynamics
  software package - quickstep: Efficient and accurate electronic structure
  calculations},\ }\href {https://doi.org/10.1063/5.0007045} {\bibfield
  {journal} {\bibinfo  {journal} {J.~Chem.~Phys.~}\ }\textbf {\bibinfo {volume}
  {152}},\ \bibinfo {pages} {194103} (\bibinfo {year} {2020})}\BibitemShut
  {NoStop}%
\bibitem [{\citenamefont {VandeVondele}\ and\ \citenamefont
  {Hutter}(2007)}]{molopt2007}%
  \BibitemOpen
  \bibfield  {author} {\bibinfo {author} {\bibfnamefont {J.}~\bibnamefont
  {VandeVondele}}\ and\ \bibinfo {author} {\bibfnamefont {J.}~\bibnamefont
  {Hutter}},\ }\bibfield  {title} {\bibinfo {title} {Gaussian basis sets for
  accurate calculations on molecular systems in gas and condensed phases},\
  }\href {https://doi.org/10.1063/1.2770708} {\bibfield  {journal} {\bibinfo
  {journal} {J.~Chem.~Phys.~}\ }\textbf {\bibinfo {volume} {127}},\ \bibinfo
  {pages} {114105} (\bibinfo {year} {2007})}\BibitemShut {NoStop}%
\bibitem [{\citenamefont {Goedecker}\ \emph {et~al.}(1996)\citenamefont
  {Goedecker}, \citenamefont {Teter},\ and\ \citenamefont {Hutter}}]{GTH1996}%
  \BibitemOpen
  \bibfield  {author} {\bibinfo {author} {\bibfnamefont {S.}~\bibnamefont
  {Goedecker}}, \bibinfo {author} {\bibfnamefont {M.}~\bibnamefont {Teter}},\
  and\ \bibinfo {author} {\bibfnamefont {J.}~\bibnamefont {Hutter}},\
  }\bibfield  {title} {\bibinfo {title} {Separable dual-space gaussian
  pseudopotentials},\ }\href {https://doi.org/10.1103/PhysRevB.54.1703}
  {\bibfield  {journal} {\bibinfo  {journal} {Phys. Rev. B}\ }\textbf {\bibinfo
  {volume} {54}},\ \bibinfo {pages} {1703} (\bibinfo {year}
  {1996})}\BibitemShut {NoStop}%
\bibitem [{\citenamefont {Hartwigsen}\ \emph {et~al.}(1998)\citenamefont
  {Hartwigsen}, \citenamefont {Goedecker},\ and\ \citenamefont
  {Hutter}}]{GTH1998}%
  \BibitemOpen
  \bibfield  {author} {\bibinfo {author} {\bibfnamefont {C.}~\bibnamefont
  {Hartwigsen}}, \bibinfo {author} {\bibfnamefont {S.}~\bibnamefont
  {Goedecker}},\ and\ \bibinfo {author} {\bibfnamefont {J.}~\bibnamefont
  {Hutter}},\ }\bibfield  {title} {\bibinfo {title} {Relativistic separable
  dual-space gaussian pseudopotentials from h to rn},\ }\href
  {https://doi.org/10.1103/PhysRevB.58.3641} {\bibfield  {journal} {\bibinfo
  {journal} {Phys. Rev. B}\ }\textbf {\bibinfo {volume} {58}},\ \bibinfo
  {pages} {3641} (\bibinfo {year} {1998})}\BibitemShut {NoStop}%
\bibitem [{\citenamefont {Krack}(2005)}]{GTH2005}%
  \BibitemOpen
  \bibfield  {author} {\bibinfo {author} {\bibfnamefont {M.}~\bibnamefont
  {Krack}},\ }\bibfield  {title} {\bibinfo {title} {Pseudopotentials for h to
  kr optimized for gradient-corrected exchange-correlation functionals},\
  }\href {https://doi.org/10.1007/s00214-005-0655-y} {\bibfield  {journal}
  {\bibinfo  {journal} {Theo. Chem. Acc.}\ }\textbf {\bibinfo {volume} {114}},\
  \bibinfo {pages} {145} (\bibinfo {year} {2005})}\BibitemShut {NoStop}%
\bibitem [{\citenamefont {Perdew}\ \emph {et~al.}(1996)\citenamefont {Perdew},
  \citenamefont {Burke},\ and\ \citenamefont {Ernzerhof}}]{pbe}%
  \BibitemOpen
  \bibfield  {author} {\bibinfo {author} {\bibfnamefont {J.~P.}\ \bibnamefont
  {Perdew}}, \bibinfo {author} {\bibfnamefont {K.}~\bibnamefont {Burke}},\ and\
  \bibinfo {author} {\bibfnamefont {M.}~\bibnamefont {Ernzerhof}},\ }\bibfield
  {title} {\bibinfo {title} {Generalized gradient approximation made simple},\
  }\href@noop {} {\bibfield  {journal} {\bibinfo  {journal}
  {Phys.~Rev.~Lett.~}\ }\textbf {\bibinfo {volume} {77}},\ \bibinfo {pages}
  {3865} (\bibinfo {year} {1996})}\BibitemShut {NoStop}%
\bibitem [{\citenamefont {Sun}\ \emph {et~al.}(2015)\citenamefont {Sun},
  \citenamefont {Ruzsinszky},\ and\ \citenamefont {Perdew}}]{sun+15prl}%
  \BibitemOpen
  \bibfield  {author} {\bibinfo {author} {\bibfnamefont {J.}~\bibnamefont
  {Sun}}, \bibinfo {author} {\bibfnamefont {A.}~\bibnamefont {Ruzsinszky}},\
  and\ \bibinfo {author} {\bibfnamefont {J.~P.}\ \bibnamefont {Perdew}},\
  }\bibfield  {title} {\bibinfo {title} {Strongly constrained and appropriately
  normed semilocal density functional},\ }\href
  {https://doi.org/10.1103/PhysRevLett.115.036402} {\bibfield  {journal}
  {\bibinfo  {journal} {Phys. Rev. Lett.}\ }\textbf {\bibinfo {volume} {115}},\
  \bibinfo {pages} {036402} (\bibinfo {year} {2015})}\BibitemShut {NoStop}%
\bibitem [{\citenamefont {Sa{\ss}nick}\ and\ \citenamefont
  {Cocchi}(2022)}]{sass-cocc22jcp}%
  \BibitemOpen
  \bibfield  {author} {\bibinfo {author} {\bibfnamefont {H.-D.}\ \bibnamefont
  {Sa{\ss}nick}}\ and\ \bibinfo {author} {\bibfnamefont {C.}~\bibnamefont
  {Cocchi}},\ }\bibfield  {title} {\bibinfo {title} {Exploring
  cesium–tellurium phase space via high-throughput calculations beyond
  semi-local density-functional theory},\ }\href@noop {} {\bibfield  {journal}
  {\bibinfo  {journal} {J.~Chem.~Phys.~}\ }\textbf {\bibinfo {volume} {156}},\
  \bibinfo {pages} {104108} (\bibinfo {year} {2022})}\BibitemShut {NoStop}%
\bibitem [{\citenamefont {Sun}\ and\ \citenamefont
  {Ceder}(2013)}]{sun-cede13ss}%
  \BibitemOpen
  \bibfield  {author} {\bibinfo {author} {\bibfnamefont {W.}~\bibnamefont
  {Sun}}\ and\ \bibinfo {author} {\bibfnamefont {G.}~\bibnamefont {Ceder}},\
  }\bibfield  {title} {\bibinfo {title} {Efficient creation and convergence of
  surface slabs},\ }\href@noop {} {\bibfield  {journal} {\bibinfo  {journal}
  {Surf.~Sci.~}\ }\textbf {\bibinfo {volume} {617}},\ \bibinfo {pages} {53}
  (\bibinfo {year} {2013})}\BibitemShut {NoStop}%
\bibitem [{\citenamefont {Jain}\ \emph {et~al.}(2013)\citenamefont {Jain},
  \citenamefont {Ong}, \citenamefont {Hautier}, \citenamefont {Chen},
  \citenamefont {Richards}, \citenamefont {Dacek}, \citenamefont {Cholia},
  \citenamefont {Gunter}, \citenamefont {Skinner}, \citenamefont {Ceder} \emph
  {et~al.}}]{jain+13aplm}%
  \BibitemOpen
  \bibfield  {author} {\bibinfo {author} {\bibfnamefont {A.}~\bibnamefont
  {Jain}}, \bibinfo {author} {\bibfnamefont {S.~P.}\ \bibnamefont {Ong}},
  \bibinfo {author} {\bibfnamefont {G.}~\bibnamefont {Hautier}}, \bibinfo
  {author} {\bibfnamefont {W.}~\bibnamefont {Chen}}, \bibinfo {author}
  {\bibfnamefont {W.~D.}\ \bibnamefont {Richards}}, \bibinfo {author}
  {\bibfnamefont {S.}~\bibnamefont {Dacek}}, \bibinfo {author} {\bibfnamefont
  {S.}~\bibnamefont {Cholia}}, \bibinfo {author} {\bibfnamefont
  {D.}~\bibnamefont {Gunter}}, \bibinfo {author} {\bibfnamefont
  {D.}~\bibnamefont {Skinner}}, \bibinfo {author} {\bibfnamefont
  {G.}~\bibnamefont {Ceder}}, \emph {et~al.},\ }\bibfield  {title} {\bibinfo
  {title} {Commentary: The materials project: A materials genome approach to
  accelerating materials innovation},\ }\href
  {https://doi.org/10.1063/1.4812323} {\bibfield  {journal} {\bibinfo
  {journal} {APL~Mater.}\ }\textbf {\bibinfo {volume} {1}},\ \bibinfo {pages}
  {011002} (\bibinfo {year} {2013})},\ \Eprint
  {https://arxiv.org/abs/https://doi.org/10.1063/1.4812323}
  {https://doi.org/10.1063/1.4812323} \BibitemShut {NoStop}%
\bibitem [{\citenamefont {Persson}(2014{\natexlab{a}})}]{CsK2Sb}%
  \BibitemOpen
  \bibfield  {author} {\bibinfo {author} {\bibfnamefont {K.}~\bibnamefont
  {Persson}},\ }\href {https://doi.org/10.17188/1276808} {\bibinfo {title}
  {Materials data on csk2sb (sg:225) by materials project}} (\bibinfo {year}
  {2014}{\natexlab{a}})\BibitemShut {NoStop}%
\bibitem [{sm()}]{sm}%
  \BibitemOpen
  \href@noop {} {}\bibinfo {note} {See Supplemental Material at [URL will be
  inserted by publisher] for additional information on the adopted methodology,
  including calculation of first- and second-order response functions, phase
  cycling, and the convergence of nuclear trajectories.}\BibitemShut {Stop}%
\bibitem [{\citenamefont {Ning}\ \emph {et~al.}(2016)\citenamefont {Ning},
  \citenamefont {Zhang}, \citenamefont {Qin}, \citenamefont {Wang},
  \citenamefont {Passerone}, \citenamefont {Ma},\ and\ \citenamefont
  {Liu}}]{ning+16ijhe}%
  \BibitemOpen
  \bibfield  {author} {\bibinfo {author} {\bibfnamefont {J.}~\bibnamefont
  {Ning}}, \bibinfo {author} {\bibfnamefont {X.}~\bibnamefont {Zhang}},
  \bibinfo {author} {\bibfnamefont {J.}~\bibnamefont {Qin}}, \bibinfo {author}
  {\bibfnamefont {L.}~\bibnamefont {Wang}}, \bibinfo {author} {\bibfnamefont
  {D.}~\bibnamefont {Passerone}}, \bibinfo {author} {\bibfnamefont
  {M.}~\bibnamefont {Ma}},\ and\ \bibinfo {author} {\bibfnamefont
  {R.}~\bibnamefont {Liu}},\ }\bibfield  {title} {\bibinfo {title} {Origin of
  distinct hydrogen absorption behavior of zr$_2$pd and zrpd$_2$},\ }\href@noop
  {} {\bibfield  {journal} {\bibinfo  {journal} {Int.~J.~Hydrog.~ Energy}\
  }\textbf {\bibinfo {volume} {41}},\ \bibinfo {pages} {1736} (\bibinfo {year}
  {2016})}\BibitemShut {NoStop}%
\bibitem [{\citenamefont {Ansari}\ \emph {et~al.}(2017)\citenamefont {Ansari},
  \citenamefont {Ulman}, \citenamefont {Camellone}, \citenamefont {Seriani},
  \citenamefont {Gebauer},\ and\ \citenamefont {Piccinin}}]{ansa+17prm}%
  \BibitemOpen
  \bibfield  {author} {\bibinfo {author} {\bibfnamefont {N.}~\bibnamefont
  {Ansari}}, \bibinfo {author} {\bibfnamefont {K.}~\bibnamefont {Ulman}},
  \bibinfo {author} {\bibfnamefont {M.~F.}\ \bibnamefont {Camellone}}, \bibinfo
  {author} {\bibfnamefont {N.}~\bibnamefont {Seriani}}, \bibinfo {author}
  {\bibfnamefont {R.}~\bibnamefont {Gebauer}},\ and\ \bibinfo {author}
  {\bibfnamefont {S.}~\bibnamefont {Piccinin}},\ }\bibfield  {title} {\bibinfo
  {title} {Hole localization in fe$_2$o$_3$ from density functional theory and
  wave-function-based methods},\ }\href@noop {} {\bibfield  {journal} {\bibinfo
   {journal} {Phys.~Rev.~Materials}\ }\textbf {\bibinfo {volume} {1}},\
  \bibinfo {pages} {035404} (\bibinfo {year} {2017})}\BibitemShut {NoStop}%
\bibitem [{\citenamefont {Lindblad}\ \emph {et~al.}(2015)\citenamefont
  {Lindblad}, \citenamefont {Jena}, \citenamefont {Philippe}, \citenamefont
  {Oscarsson}, \citenamefont {Bi}, \citenamefont {Lindblad}, \citenamefont
  {Mandal}, \citenamefont {Pal}, \citenamefont {Sarma}, \citenamefont {Karis}
  \emph {et~al.}}]{lind+1jpcc}%
  \BibitemOpen
  \bibfield  {author} {\bibinfo {author} {\bibfnamefont {R.}~\bibnamefont
  {Lindblad}}, \bibinfo {author} {\bibfnamefont {N.~K.}\ \bibnamefont {Jena}},
  \bibinfo {author} {\bibfnamefont {B.}~\bibnamefont {Philippe}}, \bibinfo
  {author} {\bibfnamefont {J.}~\bibnamefont {Oscarsson}}, \bibinfo {author}
  {\bibfnamefont {D.}~\bibnamefont {Bi}}, \bibinfo {author} {\bibfnamefont
  {A.}~\bibnamefont {Lindblad}}, \bibinfo {author} {\bibfnamefont
  {S.}~\bibnamefont {Mandal}}, \bibinfo {author} {\bibfnamefont
  {B.}~\bibnamefont {Pal}}, \bibinfo {author} {\bibfnamefont {D.~D.}\
  \bibnamefont {Sarma}}, \bibinfo {author} {\bibfnamefont {O.}~\bibnamefont
  {Karis}}, \emph {et~al.},\ }\bibfield  {title} {\bibinfo {title} {Electronic
  structure of ch$_3$nh$_3$pbx$_3$ perovskites: dependence on the halide
  moiety},\ }\href@noop {} {\bibfield  {journal} {\bibinfo  {journal}
  {J.~Phys.~Chem.~C}\ }\textbf {\bibinfo {volume} {119}},\ \bibinfo {pages}
  {1818} (\bibinfo {year} {2015})}\BibitemShut {NoStop}%
\bibitem [{\citenamefont {Persson}(2014{\natexlab{b}})}]{Cs}%
  \BibitemOpen
  \bibfield  {author} {\bibinfo {author} {\bibfnamefont {K.}~\bibnamefont
  {Persson}},\ }\href {https://doi.org/10.17188/1184811} {\bibinfo {title}
  {Materials data on cs (sg:229) by materials project}} (\bibinfo {year}
  {2014}{\natexlab{b}})\BibitemShut {NoStop}%
\bibitem [{\citenamefont {Persson}(2016)}]{K}%
  \BibitemOpen
  \bibfield  {author} {\bibinfo {author} {\bibfnamefont {K.}~\bibnamefont
  {Persson}},\ }\href {https://doi.org/10.17188/1276747} {\bibinfo {title}
  {Materials data on k (sg:229) by materials project}} (\bibinfo {year}
  {2016})\BibitemShut {NoStop}%
\bibitem [{\citenamefont {Persson}(2015)}]{Sb}%
  \BibitemOpen
  \bibfield  {author} {\bibinfo {author} {\bibfnamefont {K.}~\bibnamefont
  {Persson}},\ }\href {https://doi.org/10.17188/1187090} {\bibinfo {title}
  {Materials data on sb (sg:166) by materials project}} (\bibinfo {year}
  {2015})\BibitemShut {NoStop}%
\bibitem [{\citenamefont {Noguera}(2000)}]{nogu00jpcm}%
  \BibitemOpen
  \bibfield  {author} {\bibinfo {author} {\bibfnamefont {C.}~\bibnamefont
  {Noguera}},\ }\bibfield  {title} {\bibinfo {title} {Polar oxide surfaces},\
  }\href@noop {} {\bibfield  {journal} {\bibinfo  {journal}
  {J.~Phys.:~Condens.~Matter.~}\ }\textbf {\bibinfo {volume} {12}},\ \bibinfo
  {pages} {R367} (\bibinfo {year} {2000})}\BibitemShut {NoStop}%
\bibitem [{\citenamefont {Yue}\ \emph {et~al.}(2022)\citenamefont {Yue},
  \citenamefont {Sui}, \citenamefont {Zhao}, \citenamefont {Ni}, \citenamefont
  {Meng},\ and\ \citenamefont {Dai}}]{yue+22prb}%
  \BibitemOpen
  \bibfield  {author} {\bibinfo {author} {\bibfnamefont {T.}~\bibnamefont
  {Yue}}, \bibinfo {author} {\bibfnamefont {P.}~\bibnamefont {Sui}}, \bibinfo
  {author} {\bibfnamefont {Y.}~\bibnamefont {Zhao}}, \bibinfo {author}
  {\bibfnamefont {J.}~\bibnamefont {Ni}}, \bibinfo {author} {\bibfnamefont
  {S.}~\bibnamefont {Meng}},\ and\ \bibinfo {author} {\bibfnamefont
  {Z.}~\bibnamefont {Dai}},\ }\bibfield  {title} {\bibinfo {title} {Theoretical
  prediction of mechanics, transport, and thermoelectric properties of full
  heusler compounds na$_2$ksb and x$_2$cssb (x= k, rb)},\ }\href@noop {}
  {\bibfield  {journal} {\bibinfo  {journal} {Phys.~Rev.~B}\ }\textbf {\bibinfo
  {volume} {105}},\ \bibinfo {pages} {184304} (\bibinfo {year}
  {2022})}\BibitemShut {NoStop}%
\bibitem [{\citenamefont {Wang}\ \emph {et~al.}(2017)\citenamefont {Wang},
  \citenamefont {Pandey}, \citenamefont {Moody},\ and\ \citenamefont
  {Batista}}]{wang+17jpcc}%
  \BibitemOpen
  \bibfield  {author} {\bibinfo {author} {\bibfnamefont {G.}~\bibnamefont
  {Wang}}, \bibinfo {author} {\bibfnamefont {R.}~\bibnamefont {Pandey}},
  \bibinfo {author} {\bibfnamefont {N.~A.}\ \bibnamefont {Moody}},\ and\
  \bibinfo {author} {\bibfnamefont {E.~R.}\ \bibnamefont {Batista}},\
  }\bibfield  {title} {\bibinfo {title} {Degradation of alkali-based
  photocathodes from exposure to residual gases: a first-principles study},\
  }\href@noop {} {\bibfield  {journal} {\bibinfo  {journal} {J.~Phys.~Chem.~C}\
  }\textbf {\bibinfo {volume} {121}},\ \bibinfo {pages} {8399} (\bibinfo {year}
  {2017})}\BibitemShut {NoStop}%
\bibitem [{\citenamefont {Ghosh}\ and\ \citenamefont
  {Varma}(1978)}]{ghos-varm78jap}%
  \BibitemOpen
  \bibfield  {author} {\bibinfo {author} {\bibfnamefont {C.}~\bibnamefont
  {Ghosh}}\ and\ \bibinfo {author} {\bibfnamefont {B.}~\bibnamefont {Varma}},\
  }\bibfield  {title} {\bibinfo {title} {Preparation and study of properties of
  a few alkali antimonide photocathodes},\ }\href@noop {} {\bibfield  {journal}
  {\bibinfo  {journal} {J.~Appl.~Phys.~}\ }\textbf {\bibinfo {volume} {49}},\
  \bibinfo {pages} {4549} (\bibinfo {year} {1978})}\BibitemShut {NoStop}%
\bibitem [{\citenamefont {Bazarov}\ \emph {et~al.}(2011)\citenamefont
  {Bazarov}, \citenamefont {Cultrera}, \citenamefont {Bartnik}, \citenamefont
  {Dunham}, \citenamefont {Karkare}, \citenamefont {Li}, \citenamefont {Liu},
  \citenamefont {Maxson},\ and\ \citenamefont {Roussel}}]{baza+11apl}%
  \BibitemOpen
  \bibfield  {author} {\bibinfo {author} {\bibfnamefont {I.}~\bibnamefont
  {Bazarov}}, \bibinfo {author} {\bibfnamefont {L.}~\bibnamefont {Cultrera}},
  \bibinfo {author} {\bibfnamefont {A.}~\bibnamefont {Bartnik}}, \bibinfo
  {author} {\bibfnamefont {B.}~\bibnamefont {Dunham}}, \bibinfo {author}
  {\bibfnamefont {S.}~\bibnamefont {Karkare}}, \bibinfo {author} {\bibfnamefont
  {Y.}~\bibnamefont {Li}}, \bibinfo {author} {\bibfnamefont {X.}~\bibnamefont
  {Liu}}, \bibinfo {author} {\bibfnamefont {J.}~\bibnamefont {Maxson}},\ and\
  \bibinfo {author} {\bibfnamefont {W.}~\bibnamefont {Roussel}},\ }\bibfield
  {title} {\bibinfo {title} {Thermal emittance measurements of a cesium
  potassium antimonide photocathode},\ }\href@noop {} {\bibfield  {journal}
  {\bibinfo  {journal} {Appl.~Phys.~Lett.~}\ }\textbf {\bibinfo {volume}
  {98}},\ \bibinfo {pages} {224101} (\bibinfo {year} {2011})}\BibitemShut
  {NoStop}%
\bibitem [{\citenamefont {Xie}\ \emph {et~al.}(2016)\citenamefont {Xie},
  \citenamefont {Ben-Zvi}, \citenamefont {Rao}, \citenamefont {Xin},\ and\
  \citenamefont {Wang}}]{xie+16prab}%
  \BibitemOpen
  \bibfield  {author} {\bibinfo {author} {\bibfnamefont {H.}~\bibnamefont
  {Xie}}, \bibinfo {author} {\bibfnamefont {I.}~\bibnamefont {Ben-Zvi}},
  \bibinfo {author} {\bibfnamefont {T.}~\bibnamefont {Rao}}, \bibinfo {author}
  {\bibfnamefont {T.}~\bibnamefont {Xin}},\ and\ \bibinfo {author}
  {\bibfnamefont {E.}~\bibnamefont {Wang}},\ }\bibfield  {title} {\bibinfo
  {title} {Experimental measurements and theoretical model of the cryogenic
  performance of bialkali photocathode and characterization with monte carlo
  simulation},\ }\href@noop {} {\bibfield  {journal} {\bibinfo  {journal}
  {Phys.~Rev.~Accel.~Beams}\ }\textbf {\bibinfo {volume} {19}},\ \bibinfo
  {pages} {103401} (\bibinfo {year} {2016})}\BibitemShut {NoStop}%
\bibitem [{\citenamefont {Spicer}(1958)}]{spic58pr}%
  \BibitemOpen
  \bibfield  {author} {\bibinfo {author} {\bibfnamefont {W.~E.}\ \bibnamefont
  {Spicer}},\ }\bibfield  {title} {\bibinfo {title} {Photoemissive,
  photoconductive, and optical absorption studies of alkali-antimony
  compounds},\ }\href {https://doi.org/10.1103/PhysRev.112.114} {\bibfield
  {journal} {\bibinfo  {journal} {Phys.~Rev.~}\ }\textbf {\bibinfo {volume}
  {112}},\ \bibinfo {pages} {114} (\bibinfo {year} {1958})}\BibitemShut
  {NoStop}%
\bibitem [{\citenamefont {Karkare}\ and\ \citenamefont
  {Bazarov}(2015)}]{kark-baza15prap}%
  \BibitemOpen
  \bibfield  {author} {\bibinfo {author} {\bibfnamefont {S.}~\bibnamefont
  {Karkare}}\ and\ \bibinfo {author} {\bibfnamefont {I.}~\bibnamefont
  {Bazarov}},\ }\bibfield  {title} {\bibinfo {title} {Effects of surface
  nonuniformities on the mean transverse energy from photocathodes},\
  }\href@noop {} {\bibfield  {journal} {\bibinfo  {journal} {Phys.~Rev.~Appl.}\
  }\textbf {\bibinfo {volume} {4}},\ \bibinfo {pages} {024015} (\bibinfo {year}
  {2015})}\BibitemShut {NoStop}%
\end{thebibliography}

%

\end{document}